\documentclass[12pt]{article}
\usepackage[]{geometry}
\usepackage{mathtools}
\usepackage{bigstrut}
\usepackage{mathrsfs}
\usepackage{setspace}
\usepackage{amsfonts}
\usepackage{amssymb}
\usepackage{amsmath}
\usepackage{enumitem}
\usepackage{latexsym}
\usepackage{multirow}
\usepackage{booktabs}
\usepackage[authoryear]{natbib}

\usepackage[pdftex]{color,graphicx}
\definecolor{lgray}{gray}{0.70}
\newcommand{\function}[2]{#1\left(#2\right)}
\newcommand{\p}[1]{\mathbf{P}\left(#1\right)}

\newcommand{\e}[1]{\mathbf{E}\left(#1\right)}
\newcommand{\med}[1]{\mathbf{med}\left(#1\right)}
\newcommand{\var}[1]{\mathbf{Var}\left(#1\right)}
\newenvironment{proof}{\textbf{Proof: }}{$\hfill \square \\$}

\newtheorem{lemma}{Lemma}[section]
\newtheorem{prop}{Proposition}[section]
\newtheorem{thm}{Theorem}[section]
\newtheorem{cor}{Corollary}[section]

\newtheorem{remark}[prop]{Remark}
\newcommand{\ind}[1]{\mathbf{1}_{#1}}
\newcommand{\cas}{\stackrel{a.s.}{\longrightarrow}}
\newcommand{\cip}{\stackrel{\mathbf{P}}{\longrightarrow}}

\newcommand{\argmax}{\operatornamewithlimits{\textrm{argmax}}}

\newcommand{\lsup}{\operatornamewithlimits{\overline{\lim}}}

\newcommand{\R}{\mathbb{R}}
\usepackage[colorlinks,citecolor=blue,urlcolor=blue,linkcolor=red,naturalnames=true]{hyperref}

\begin{document}
\title{A consistent bootstrap procedure for the maximum score estimator}
\author{\begin{tabular}{c}
Rohit Kumar Patra, Emilio Seijo, and Bodhisattva Sen\footnote{Corresponding author. Tel.: +1 212 851 2149; fax: +1 212 851 2164. E-mail address: bodhi@stat.columbia.edu.} \footnote{Research partially supported by NSF grant DMS-1150435 (CAREER)}\\
Department of Statistics, Columbia University
\end{tabular}}
\maketitle
\begin{abstract}
\noindent In this paper we propose a new model-based smoothed bootstrap procedure for making inference on the maximum score estimator of~\cite{man75,man85} and prove its consistency. We provide a set of sufficient conditions for the consistency of any bootstrap procedure in this problem. We compare the finite sample performance of different bootstrap procedures through simulations studies. The results indicate that our proposed smoothed bootstrap outperforms other bootstrap schemes, including the $m$-out-of-$n$ bootstrap. Additionally, we prove a convergence theorem for triangular arrays of random variables arising from binary choice models, which may be of independent interest. \newline


\end{abstract}

{\small }
\section{Introduction}\label{s1}
Consider a (latent-variable) binary response model of the form 
\begin{equation*}
Y = \ind{\beta_0^\top X + U\geq 0},
\end{equation*}
where $\mathbf{1}$ is the indicator function, $X$ is an $\mathbb{R}^d$-valued continuous random vector of explanatory variables, $U$ is an unobserved random variable and $\beta_0\in\mathbb{R}^d$ is an unknown vector with $|\beta_0|=1$ ($|\cdot|$ denotes the Euclidean norm in $\mathbb{R}^d$). The parameter of interest is $\beta_0$. If the conditional distribution of $U$ given $X$ is known up to a finite set of parameters, maximum likelihood techniques can be used for estimation, among other methods; see, e.g., \cite{mc74}. The parametric assumption on $U$ may be relaxed in several ways. For instance,  if $U$ and $X$ are independent or if the distribution of $U$ depends on $X$ only through the index $\beta_0^\top X$, the semiparametric estimators of \cite{h87}, \cite{hh96}, \cite{pss89}, and \cite{s93} can be used; also see \cite{c83}. The maximum score estimator considered by \cite{man75} permits the distribution of $U$ to depend on $X$ in an unknown and very general way (heteroscedasticity of unknown form). The model replaced parametric assumptions on the error disturbance $U$ with a conditional median restriction, i.e., $\med{U|X}=0$, where $\med{U|X}$ represents the conditional median of $U$ given $X$. Given $n$ observations $(X_1,Y_1),\ldots,(X_n,Y_n)$ from such a model, \cite{man75} defined a {\it maximum score estimator} as any maximizer of the objective function $$\sum_{i=1}^n \left(Y_i - \frac{1}{2}\right)\ind{\beta^\top X_i \geq 0}$$ over the unit sphere in $\mathbb{R}^d$.

 The asymptotics for the maximum score estimator are well-known. Under some regularity conditions, the estimator was shown to be strongly consistent in \cite{man85} and its asymptotic distribution was derived in \cite{kipo} (also see \cite{c87}). Even though the maximum score estimator is the most general estimator available for the binary response model considered here, the complicated nature of its limit law (which depends, among other parameters, on the conditional distribution of $U$ given $X$ for values of $X$ on the hyperplane $\{x\in\mathbb{R}^d:\beta_0^\top x = 0\}$) and the fact that it exhibits nonstandard asymptotics (cube-root rate of convergence) have made it difficult to do inference for the estimator under complete generality. 

As an alternative, \cite{ho1992} proposed the smoothed maximum score estimator. Although this estimator is asymptotically normally distributed under certain assumptions (after proper centering and scaling) and the classical bootstrap can be used for inference (see \cite{ho2002}; also see \cite{DeWoutersen11} for extensions to certain dependence structures), it has a number of drawbacks: it requires stronger assumptions on the model for the asymptotic results to hold, the smoothing of the score function induces bias which can be  problematic to deal with, and the plug-in methods (see~\cite{ho1992, ho2002}) used to correct for this bias are not effective when the model is heteroscedastic or multimodal (see~\cite{kotlyarova09}).


This motivates us to study the maximum score estimator and investigate the performance of bootstrap ---  a natural alternative for inference in such nonstandard problems. Bootstrap methods avoid the problem of estimating nuisance parameters and are generally reliable in problems with $n^{-1/2}$ convergence rate and  Gaussian limiting distributions; see \cite{bf81}, \cite{s81}, \cite{st95} and its references. Unfortunately, the classical bootstrap (drawing $n$ observations with replacement from the original data) is {\it inconsistent} for the maximum score estimator as shown in \cite{ah05}. In fact, the classical bootstrap can behave quite erratically in cube-root convergence problems. For instance, it was shown in \cite{sebawo} that for the Grenander estimator (the nonparametric maximum likelihood estimator of a non-increasing density on $[0,\infty)$), a prototypical example of cube-root asymptotics, the bootstrap estimator is not only inconsistent but has no weak limit in probability. This stronger result should also hold for the maximum score estimator. These findings contradict some of the results of \cite{ah05} (especially Theorem 4 and the conclusions of Section 4 of that paper) where it is claimed that for some single-parameter estimators a simple method for inference based on the classical bootstrap can be developed in spite of its inconsistency.

Thus, in order to apply the bootstrap to this problem some modifications of the classical approach are required. Two variants of the classical bootstrap that can be applied in this situation are the so-called $m$-out-of-$n$ bootstrap and subsampling. The performance of subsampling for inference on the maximum score estimator has been studied in \cite{deromi01}. The consistency of the $m$-out-of-$n$ bootstrap can be deduced from the results in \cite{LP06}. Despite their simplicity, the reliability of both methods depends crucially on the size of the subsample (the $m$ in the $m$-out-of-$n$ bootstrap and the block size in subsampling) and a proper choice of this tuning parameter is difficult; see Section 4 of \cite{LP06} for a brief discussion on this.  Thus, it would be desirable to have other alternatives --- more automated and consistent bootstrap procedures  --- for inference in the general setting of the binary choice model of Manski.

In this paper we propose a model-based smoothed bootstrap procedure (i.e., a method that uses the model setup and assumptions explicitly to construct the bootstrap scheme; see Section \ref{s2s2s3} for the details) that provides an alternative to subsampling and the $m$-out-of-$n$ bootstrap. We prove that the procedure is consistent for the maximum score estimator. In doing so, we state and prove a general convergence theorem for triangular arrays of random variables coming from binary choice models that can be used to verify the consistency of any bootstrap scheme in this setup. 
 We derive our results in greater generality\footnote{\noindent We do not need to assume that the coefficient corresponding to a particular covariate is non-zero.} than most authors by assuming that $\beta_0$ belongs to the unit sphere in $\mathbb{R}^d$ as opposed to fixing its first co-ordinate to be 1 (as in \cite{ah05}).  To make the final results more accessible we express them in terms of integrals with respect to the Lebesgue measure as opposed to surface measures, as in \cite{kipo}. 
 We run simulation experiments to compare the finite sample performance of different bootstrap procedures. Our results indicate that the proposed smoothed bootstrap method (see Section \ref{s2s2s3}) outperforms all the others. Even though the proposed bootstrap scheme involves the choice of tuning parameters, they are easy to tune --- smoothing bandwidths that fit the data well are to be preferred.
 
 To the best of our knowledge, this paper is the first attempt to understand the behavior of model-based bootstrap procedures under the very general heteroscedasticity assumptions for the maximum score estimator.

Our exposition is organized as follows: In Section \ref{s2s1} we introduce the model and our assumptions. In Section \ref{snew3} we  propose the smoothed bootstrap procedure for the maximum score estimator and discuss its consistency. 
 We study and compare the finite sample performance of the different bootstrap schemes in Section \ref{s5} through simulation experiments.  In Section \ref{s3} we state a general convergence theorem for triangular arrays of random variables coming from binary choice models (see Theorem \ref{t1}) which is useful in proving the consistency of our proposed bootstrap scheme (given in Section~\ref{sec:mainproof}). 
 Section~\ref{app:lemma_proof} gives the proofs of the results in Section~\ref{s3}. Appendix~\ref{app1} contains some auxiliary results and some technical details omitted from the main text. In Appendix \ref{Disc} we provide a necessary and sufficient condition for the existence of the latent variable structure in a binary choice model, that may be of independent interest.

\section{The model}\label{s2s1}
We start by introducing some notation. For a signed Borel measure $\mu$ on some metric space $\texttt{X}$ and a Borel measurable function $f:\texttt{X}\rightarrow\mathbb{R}$ which is either integrable or nonnegative we will use the notation $\mu(f) := \int f d\mu$. If $\mathcal{G}$ is a class of such functions on $\texttt{X}$ we write $\|\mu\|_\mathcal{G}:=\sup\{|\mu(f)|:f\in\mathcal{G}\}$.  We will also make use of the sup-norm notation, i.e., for functions $g:\texttt{X}\rightarrow\mathbb{R}^d$, $G:\texttt{X}\rightarrow\mathbb{R}^{d\times d}$ we write $\|g\|_\texttt{X} := \sup\{|g(x)|:x\in\texttt{X}\}$ and $\|G\|_\texttt{X} := \sup\{\|G(x)\|_2 :x\in\texttt{X}\}$, where $|\cdot|$ stands for the usual Euclidean norm and $\|\cdot\|_2$ denotes the matrix $\mathbb{L}_2$-norm on the space $\mathbb{R}^{d\times d}$ of all $d \times d$ real matrices (see \cite{meyerma}, page 281).  For a  differentiable  function $f: \R^d  \rightarrow \R$ we write $\nabla f(x) :=\partial f/\partial x$ for its gradient at $x$.  We will regard the elements of Euclidean spaces as column vectors.  For two real numbers $a$ and $b,$ we write $ a \land b :=\min (a, b)$ and $a \lor b:= \max (a,b).$

Consider a Borel probability measure $ \mathbb{\tilde{P}}$ on $\mathbb{R}^{d+1}$, $d\geq 2$, such that if $(X,U)\sim \mathbb{\tilde{P}}$ then $X$ takes values in a  closed, convex region $\mathfrak{X}\subset\mathbb{R}^d$ with $\mathfrak{X}^\circ\neq\emptyset$ (here $\mathfrak{X}^\circ$ denotes the interior of the set $\mathfrak{X}$) and $U$ is a real-valued random variable that satisfies $\med{U|X}=0$ almost surely (a.s.), where $\med{\cdot}$ represents the median. We only observe $(X,Y) \sim \mathbb{P}$ where 
\begin{equation}\label{eq:Mdl}
Y:=\ind{\beta_0^\top X+U\geq 0}
\end{equation} 
for some $\beta_0\in\mathcal{S}^{d-1}$ ($\mathcal{S}^{d-1}$ is the unit sphere in $\mathbb{R}^d$ with respect to the Euclidean~norm). Throughout the paper we assume the following conditions on the distribution $\mathbb{P}$:

\begin{enumerate}  [label=\bfseries (C\arabic*)]
	\item $\mathfrak{X}$ is a convex and compact subset of $\R^d$. \label{asum_c1}
	\item Under $\mathbb{P}$, $X$ has a continuous distribution with a strictly positive and continuously differentiable density $p$ on $\mathfrak{X}^\circ$. We also assume that $\nabla p$ is integrable (with respect to the Lebesgue measure) over $\mathfrak{X}$. Let $F$ denote the distribution of $X$ under $\mathbb{P}$, i.e., $F(A):=\p{X\in A}$, for $A \subset \R^d$ Borel.\label{asum_c2}
	
	\item Define 
	\begin{equation}\label{eq:kappa}
\kappa(x):= \mathbb{P}\left( Y = 1|X=x\right) = \mathbb{\tilde{P}}\left(\beta_0^\top X + U \geq 0 | X=x\right).
\end{equation}
We assume that $\kappa$ is continuously differentiable on $\mathfrak{X}^\circ$, the set $\{x\in\mathfrak{X}^\circ: \nabla\kappa(x)^\top \beta_0>0\}$ intersects the hyperplane $\{x\in\mathbb{R}^d: \beta_0^\top x = 0\}$, and that $\displaystyle \int|\nabla\kappa(x)|xx^\top p(x) dx$ is well-defined.\label{asum_c3}
\end{enumerate}

Given observations $(X_1,Y_1),\ldots,(X_n,Y_n)$ from such a model, we wish to estimate $\beta_0\in\mathcal{S}^{d-1}$. A maximum score estimator of $\beta_0$ is any element $\hat{\beta}_n\in\mathcal{S}^{d-1}$ that satisfies:
\begin{equation}\label{ec1}
\hat{\beta}_n := \argmax_{\beta\in\mathcal{S}^{d-1}}\left\{ \frac{1}{n}\sum_{i=1}^n \left(Y_i - \frac{1}{2}\right)\ind{\beta^\top X_i\geq 0}\right\}.
\end{equation}
Note that there may be many elements of $\mathcal{S}^{d-1}$ that satisfy $(\ref{ec1})$. We will focus on {\it measurable selections} of maximum score estimators, i.e., we will assume that we can compute the estimator in such a way that $\hat{\beta}_n$ is measurable (this is justified in view of the {\it measurable selection theorem}, see Chapter 8 of \cite{aufra2009}). We make this assumption to avoid the use of outer probabilities.


Our assumptions \ref{asum_c1}--\ref{asum_c2} on $\mathbb{P}$ and the continuous differentiability of $\kappa$ imply that $\Gamma(\beta)$,  defined as 
\begin{equation}\label{eq:DefGamma}
\Gamma(\beta):= \mathbb{P}\left[\left(Y-\frac{1}{2}\right)\ind{\beta^\top X\geq 0}\right]
\end{equation}
is twice continuously differentiable in a neighborhood of $\beta_0$ (see Lemma \ref{l3}). Moreover, condition~\ref{asum_c3} implies that the Hessian matrix $\nabla^2\Gamma(\beta_0)$ is non-positive definite on an open neighborhood $U\subset\mathbb{R}^{d}$ of $\beta_0$; see Lemma~\ref{l3}. Our regularity conditions \ref{asum_c1}--\ref{asum_c3} are equivalent to those in Example 6.4 of \cite{kipo} and imply those in \cite{man85}.  Hence, a consequence of Lemmas 2 and 3 in \cite{man85} is that $\beta_0$ is identifiable and is the unique maximizer of the process $\Gamma(\beta)$ where $\beta \in \mathcal{S}^{d-1}$. Similarly, Theorem 1 in the same paper implies that if $(\hat{\beta}_n)_{n=1}^\infty$ is any sequence of maximum score estimators, we have $\hat{\beta}_n\cas\beta_0$.

\section{Bootstrap and consistency}\label{snew3}
\subsection{Smoothed bootstrap}\label{s2s2s3}
In this sub-section we propose a smoothed bootstrap procedure for constructing confidence regions for $\beta_0$. Observe that $Y$ given $X=x$ follows a Bernoulli distribution with probability of ``success'' $\kappa(x)$, i.e., $Y|X = x \sim$ Bernoulli$(\kappa(x))$, where $\kappa$ is defined in~\eqref{eq:kappa}. Our bootstrap procedure is model-based and it exploits the above relationship between $Y$ and $X$ using a nonparametric estimator of $\kappa$.  The smoothed bootstrap procedure can be described as follows:

\begin{enumerate}[label=(\roman*)]
\item Choose an appropriate nonparametric smoothing procedure (e.g., kernel density estimation) to construct a density estimator $\hat p_n$ of $p$ using  $X_1, \ldots,X_n$. 

    \item Use $(X_1,Y_1,), \ldots, (Y_n,X_n)$ to find a smooth estimator $\hat{\kappa}_n$ of $\kappa$ (e.g., using kernel regression). 
    
	\item Sample $(X_{n,1}^*, Y_{n,1}^*), \ldots, (X_{n,n}^*, Y_{n,n}^*)$ i.i.d.~$\widehat{\mathbb{Q}}_n$ (conditional on the data), where $(X,Y) \sim \widehat{\mathbb{Q}}_n$ if and only if $X \sim \hat p_n$ and $Y|X = x \sim$ Bernoulli$(\hat \kappa_n(x))$.

    \item Let $\hat \beta_n^*$ be any maximizer of
    \begin{equation*}\nonumber
    S_n^*(\beta):=\frac{1}{n}\sum_{i=1}^n \left(Y_{n,i}^* -\frac{1}{2}\right)\ind{\beta^\top X_{n,i}^*\geq 0}.
    \end{equation*}
    
    	\item Compute 
\begin{equation}\label{eq:BootsMSE}	
	\tilde \beta_n = \argmax_{\beta \in \mathcal{S}^{d-1}} \int_{\beta^\top x \ge 0} \left\{\hat \kappa_n(x) - \frac{1}{2} \right\} \hat p_n(x) dx.
\end{equation}
	
\end{enumerate}
%
Let $\mathbb{G}_n$ be the distribution of the (normalized and centered) maximum score estimator, i.e.,
\begin{equation}\label{eq:Delta}
 \Delta_n := n^{1/3} (\hat \beta_n - \beta_0) \sim \mathbb{G}_n.
\end{equation} 
\cite{kipo} showed that, under conditions \ref{asum_c1}--\ref{asum_c3}, $\Delta_n$ converges in distribution. Let $\mathbb{G}$ denote the distribution of this limit. Thus, 
\begin{equation*}\label{eq:pollard_conv}
\rho(\mathbb{G}_n,\mathbb{G}) \rightarrow 0,
\end{equation*}
 where $\rho$ is the Prokhorov metric or any other metric metrizing weak convergence of probability measures.  Moreover, let  $\hat{\mathbb{G}}_n$  be the conditional distribution of 
\begin{equation}\label{eq:DeltaStar}
\Delta_n^* := n^{1/3}(\hat \beta_n^* - \tilde{\beta}_n)
\end{equation} 
 given  the data, i.e., for any Borel set $A \subset \R^d$, $\hat{\mathbb{G}}_n(A) = \p{\Delta_n^* \in A\big|\function{\sigma}{\left(X_n,Y_n\right)_{n=1}^\infty}}$. We will approximate  $\mathbb{G}_n$ by $\hat{\mathbb{G}}_n$, and use this to build confidence sets for $\beta_0$. In Section~\ref{s4}, we  will show that the smoothed bootstrap scheme is weakly consistent, i.e., 
\begin{equation*}
\rho(\mathbb{G}_n, \hat{\mathbb{G}}_n) \cip 0.
\end{equation*}
 It was shown in \cite{ah05} that the bootstrap procedure based on sampling with replacement from the data $(X_1,Y_1),\ldots, (X_n,Y_n)$ (the classical bootstrap) is inconsistent. 
 
Steps (i)---(v) deserve comments. We start with (i). It will be seen in Theorem \ref{t1} that the asymptotic distribution of $\Delta_n$ depends on the behavior of $F$, the distribution of $X$ under $\mathbb{P}$, around the hyperplane $\mathcal{H}:=\{x\in\mathbb{R}^d:\ \beta_0^\top x=0\}$. As the empirical distribution is discrete, a smooth approximation to $F$ might yield a better finite sample approximation to the local behavior around $\mathcal{H}$. Indeed our simulation studies clearly illustrate this point (see Section \ref{s5}). We can use any nonparametric density estimation method to estimate $p$. In our simulation studies, we  use  the ``product kernel function'' constructed from a product of $d$ univariate kernel functions and estimate $p$ by
\begin{equation}\label{NPdensity}
	\hat p_n(x) = \frac{1}{n h_{n,1} \cdots h_{n,d}} \sum_{i=1}^n K\left( \frac{x - X_i}{h_n} \right), \qquad \mbox{ for } x \in \R^d,
\end{equation} 
where $h_n = (h_{n,1}, \ldots, h_{n,d})$, $K\left( \frac{x - X_i}{h_n} \right) := k\left( \frac{x_1 - X_{i,1}}{h_{n,1}} \right)\times \ldots \times k\left( \frac{x_d - X_{i,d}}{h_{n,d}} \right)$, and $k(\cdot)$ is the density function of a symmetric random variable with finite variance; see~\cite{EM05} and the references therein for the consistency of kernel-type function estimators. 

As noted after \eqref{eq:kappa}, $\kappa$ plays a central role in determining the joint distribution of $(X,Y)$ and in the absence of any prior knowledge on the conditional distribution function of $Y$ given $X$ we can estimate it nonparametrically using the Nadaraya-Watson estimator
\begin{equation}\label{eq:NPkappa}
	\hat \kappa_n(x) = \frac{\sum_{i=1}^n Y_i K((x - X_i)/h_n)}{\sum_{i=1}^n K((x - X_i)/h_n)},
\end{equation}
where $h_n \in \R^d$ is the bandwidth vector and $K:\R^d \to \R$ is the product kernel. A huge literature has been developed on the consistency of the Nadaraya-Watson estimator; see e.g., \cite{NPLi11} and the references therein. 

In (iii), we generate the bootstrap sample from the estimated joint distribution of $(X,Y)$. Note that our approach is completely nonparametric and allows us to model any kind of dependence between $X$ and $Y$. The maximum score estimator from the bootstrap sample is computed in (iv).

Our bootstrap  procedure does not necessarily reflect the {\it latent variable} structure in \eqref{eq:Mdl}; see Appendix~\ref{Disc}  for a detailed discussion of this and a lemma discussing a necessary and sufficient condition for the existence of the latent variable structure. Therefore, $\hat \beta_n$ is not guaranteed to be the maximum score estimator for the sampling distribution at the bootstrap stage. For the bootstrap scheme to be consistent we need to change the centering of our bootstrap estimator from $\hat \beta_n$ to $\tilde \beta_n$, the maximum score estimator obtained from the smoothed joint distribution of $(X,Y)$. This is done in (v).

\begin{remark} \label{rem:kappa_bayes}
In the above smoothed bootstrap scheme we generate i.i.d.~samples from the joint distribution of $(X,Y)$ by first drawing $X$ from its marginal distribution and then generating $Y$ from the conditional distribution of $Y|X$. A natural alternative is to draw $Y \sim$ Bernoulli$(\pi)$ first (where $\pi := \mathbb{P}(Y = 1)$) and then to generate $X$ from the conditional distribution of $X|Y$. In this approach, we need to estimate the conditional density of $X$ given $Y=0$ ($f_{X|Y=0}$) and $Y=1$ ($f_{X|Y=1}$) and $\pi$. Note that $\kappa$ and the conditional densities are related as 
\begin{equation} \label{def:kappa_2}
\kappa(x) = \frac{\pi f_{X|Y=1}(x)}{ (1-\pi) f_{X|Y=0}(x) + \pi f_{X|Y=1}(x)}.
\end{equation}
A natural estimator of $\pi$, $\hat\pi_n$, can be the relative frequency of $Y = 1$ in the observed data. Further, $f_{X|Y=0}$ and $f_{X|Y=1}$ can be estimated using standard kernel density estimation procedures after partitioning the data based on the values of $Y$.

\end{remark}


\begin{remark}
Note that $\hat{\kappa}_n$  in the smoothed bootstrap procedure, as described in Section~\ref{s2s2s3}, does not necessarily satisfy the inequality $(\hat{\beta}_n^\top x)(\hat{\kappa}_n(x)-1/2)\geq 0$ for all $x\in\mathfrak{X}$. Thus the smoothed bootstrap procedure does not strictly mimic the latent variable structure in the model. However, it must be noted that the referred inequality will be satisfied asymptotically for all $x$ outside the hyperplane $\left\{x\in\mathbb{R}^d: \beta_0^\top  x=0\right\}$ whenever $\hat{\beta}_n$ and $\hat{\kappa}_n$ are consistent. 
\end{remark}

\subsection{Consistency of smoothed bootstrap}\label{s4}
In this sub-section we study the consistency of the smoothed bootstrap procedure proposed in the previous sub-section. The classical bootstrap scheme is known to be {\it inconsistent} for the maximum score estimator; see \cite{ah05}. The consistency of subsampling and the $m$-out-of-$n$ bootstrap in this problem can be deduced from the results in  \cite{deromi01} and \cite{LP06}, respectively. However, finite sample performance of  both subsampling and the $m$-out-of-$n$ bootstrap depend crucially on the choice of the block size ($m$), and the choice of a proper $m$ is very difficult. Moreover, different choices of $m$ lead to very different results. In contrast, the tuning parameters involved in the model based smoothed bootstrap procedure can be easily calibrated --- smoothing bandwidths that fit the given data well are to be preferred. 

We recall the notation and definitions established in Section \ref{s2s1}. We will denote by $\mathscr{Z}=\function{\sigma}{\left(X_n,Y_n\right)_{n=1}^\infty}$ the $\sigma$-algebra generated by the sequence $\left(X_n,Y_n\right)_{n=1}^\infty$ with $\left(X_n,Y_n\right)_{n=1}^\infty\stackrel{i.i.d.}{\sim}\mathbb{P}$. Let $\widehat{\mathbb{Q}}_n$ be the probability measure on $\mathbb{R}^{d+1}$ such that $(X,Y) \sim \widehat{\mathbb{Q}}_n$ if and only if 
\begin{equation*}\label{eq:SmBoots}
X \sim \hat p_n \;\;\;\; \mbox{ and } \;\;\;\;Y|X = x \sim \mbox{Bernoulli}(\hat \kappa_n(x)),
\end{equation*} 
where $\hat p_n$ and $\hat \kappa_n$ are estimators of $p$ and $\kappa$ respectively, and may be defined as in~\eqref{NPdensity} and~\eqref{eq:NPkappa}. We can regard the bootstrap samples as $(X_{n,1}^*,Y_{n,1}^*),\ldots,(X_{n,n}^*,Y_{n,n}^*)\stackrel{i.i.d.}{\sim}\widehat{\mathbb{Q}}_n$. 

 Recall that  $\mathbb{G}_n$ denotes  the distribution of  $n^{1/3} (\hat\beta_n-\beta_0)$ and $\rho(\mathbb{G}_n,\mathbb{G})\rightarrow 0.$ Moreover, $\hat{\mathbb{G}}_n$ denotes the conditional distribution of $n^{1/3}(\hat \beta_n^* - \tilde{\beta}_n)$, given the data, where $\tilde{\beta}_n$ is defined in \eqref{eq:BootsMSE}.   Thus, a necessary and sufficient condition for the smoothed bootstrap procedure to be weakly consistent is 
\begin{equation}\label{eq:BootsCons}
\rho( \hat{\mathbb{G}}_n, \mathbb{G}) \cip 0.
\end{equation} 
In  the following theorem, we give sufficient conditions for the smoothed bootstrap procedure proposed to be consistent.
\begin{thm}[\textbf{Main Theorem}] \label{t3}
Consider the smoothed bootstrap scheme described in Section \ref{s2s2s3} and assume that assumptions \ref{asum_c1}--\ref{asum_c3} hold. Furthermore,  assume that the following conditions hold:
\begin{enumerate}[label=\bfseries (S\arabic*)]	
	\item  The sequence $(\hat p_n)_{n=1}^\infty$ of densities is such that $p_n$ is continously differentiable on $\mathfrak{X}^\circ$, $\nabla p_n$ is integrable  (with respect to the Lebesgue measure) over $\mathfrak{X}$, and  $\|\hat{p}_n - p\|_\mathfrak{X} = o_\mathbf{P}(n^{-1/3})$. \label{smoth_1}
	\item  $\hat{\kappa}_n$ converges to $\kappa$ uniformly on compact subsets of $\mathfrak{X}^\circ$ w.p.~1.\label{smoth_2}
	\item  For any compact set $\texttt{X} \subset \mathfrak{X}^\circ$, $\| \nabla \hat \kappa_n - \nabla \kappa \|_{\texttt{X}} \rightarrow 0$ a.s.~and $\| \nabla \hat \kappa_n - \nabla \kappa\|_{\mathfrak{X}} = O_\mathbf{P}(1)$. \label{smoth_3}
\end{enumerate}
\vspace{-.3in}
Then, the smoothed bootstrap procedure is weakly consistent, i.e., the conditional distribution of $\Delta_n^*$, given $(X_1,Y_1),\ldots,(X_n,Y_n)$, converges to $\mathbb{G}$  in probability $($see \eqref{eq:BootsCons}$)$.
\end{thm}

The proof of Theorem \ref{t3} is  involved and is given in Section~\ref{sec:mainproof}. The proof uses results from Section~\ref{s3} where we give a convergence theorem for the maximum score estimator for triangular arrays of random variables arising from the binary choice model.


 Conditions \ref{smoth_1}--\ref{smoth_3} deserve comments. In the following we discuss the existence of $\hat{p}_n$ and $\hat{\kappa}_n$ satisfying conditions \ref{smoth_1}--\ref{smoth_3}. If we use kernel density estimation techniques to construct $\hat p_n$, then Theorem 1 and Corollary 1 of~\cite{EM05} give very general conditions on the uniform in bandwidth consistency of kernel-type function estimators. In particular, they imply that \ref{smoth_1} holds if $p$ is sufficiently smooth. According to~\cite{stone1982} the optimal and achievable rate of convergence for estimating $p$ nonparametrically is $n^{-\frac{r}{2r+d}}$ if $p$ is $r$ times continuously differentiable over $\mathfrak{X}$. For the Nadaraya-Watson estimator Theorem 2 and Corollary 2 of \cite{EM05} gives numerous results on the uniform convergence of $\hat \kappa_n$ which, in particular, shows that \ref{smoth_2} holds. The first condition in \ref{smoth_3} on the uniform convergence on compacts (in the interior of the support of $X$) of $\nabla \hat \kappa_n$ holds for the Nadaraya-Watson estimator defined in \eqref{eq:NPkappa}; see \cite{Blondin07}. The second condition in \ref{smoth_3} can also be shown to hold under appropriate conditions on the smoothing bandwidth and kernel if $p$ is strictly positive on $\mathfrak{X}$.
 
\begin{remark}\label{rem:AltSmBts2}
Recall the alternative data generating mechanism described in Remark~\ref{rem:kappa_bayes}. Let $\hat \kappa_n$ now be the estimator based on~\eqref{def:kappa_2}, where we use plug-in kernel density estimators of $f_{X|Y=0}$ and $f_{X|Y=1}$, and estimate $\pi$ by the sample mean of $Y$. Then, Theorem 1 and Corollary 1 of~\cite{EM05} (note that $p(x)$ is bounded away from zero on $\mathfrak{X}$) can be used to show that $\hat p_n$ and $\hat{\kappa}_n$ satisfy conditions \ref{smoth_1}--\ref{smoth_3}. Hence, by Theorem~\ref{t3}, this smoothed bootstrap approach can also be shown to be consistent. 
\end{remark}
\section{Simulation experiments}\label{s5}
  In this section we illustrate the finite sample performance of our proposed smoothed bootstrap, the classical bootstrap, and the $m$-out-of-$n$ bootstrap through simulation experiments.
 Let $\{ (X_1^*,Y_1^*), \ldots, (X_m^*,Y_m^*)\}$ be $m$ samples drawn randomly with replacement from $ \{ (X_1,Y_1), \ldots, (X_n,Y_n)\}$. The $m$-out-of-$n$ bootstrap estimates $\mathbb{G}_n$ (see~\eqref{eq:Delta}) by the distribution of $m^{1/3} (\check{\beta}_m -\hat{\beta}_n),$ where $$\check{\beta}_m :=  \argmax_{\beta\in\mathcal{S}^{d-1}}\left\{ \frac{1}{m}\sum_{i=1}^m \left(Y_i^* - \frac{1}{2}\right)\ind{\beta^\top X_i^*\geq 0}\right\}.$$ \cite{LP06} prove that such a bootstrap procedure is weakly consistent for the binary response model considered in this paper if $m = o(n)$ and $m \to \infty$, as $n \to \infty$. However finite sample performance of $m$-out-of-$n$ bootstrap relies heavily on the choice of $m$, and a proper choice is difficult. Also, most data driven choices for $m$ are computationally very expensive. For a comprehensive overview of $m$-out-of-$n$ bootstrap methods and discussion on the choice of $m$ see \cite{BGZ97} and \cite{bickel2008choice}.


%
 In our simulation study we take  $(X,U)\sim\mathbb{\tilde P},$ where $ \mathbb{\tilde P}$ is a distribution on $\mathbb{R}^{d+1}$ and  fix  $\beta_0=\frac{1}{\sqrt{d}}(1,\ldots,1)^\top.$ For  $ \mathbb{\tilde P}$ to satisfy our model assumptions, we let  $$U|X\sim N\left(0,\frac{1}{(1+|X|^2)^{2}}\right) , \quad X\sim \mbox{Uniform}([-1,1]^d),  \quad  \mbox{ and} \quad Y = \ind{\beta_0^\top X + U\geq 0}.$$ Thus, in this case $\kappa(x)=1-\Phi(-\beta_0^\top x(1+|x|^2))$, which is, of course, infinitely differentiable. Consequently, according to \cite{stone1982}, the optimal (achievable) rates of convergence to estimate $\kappa$ nonparametrically are faster than those required in (ii) of Theorem \ref{t3}. To compute the estimator $\hat{\kappa}_n$ of $\kappa$ we have chosen to use the Nadaraya-Watson estimator with a Gaussian kernel and a bandwidth given by Scott's normal reference rule (see \cite{sco}, page 152). 
 To sample $X_{n,i}^*$, we first sample randomly with replacement from $X_1, \ldots, X_n$ and then add a $d$-dimensional independent mean zero Gaussian random variable with a diagonal variance-covariance  matrix $\mathcal{D}$, where diag$(\mathcal{D})=(h^2_{n1},\ldots, h^2_{nd})$ and $(h_{n1},\ldots, h_{nd})$  is the bandwidth vector given by Scott's normal reference rule for kernel density estimation.  Note that, this is equivalent to sampling from the kernel density estimate with Gaussian kernels and bandwidth given by Scott's normal reference rule. We would like to point out that our selection of the smoothing parameters are not optimal in any sense and could be improved by applying data-driven selection methods, such as cross-validation (e.g., see Chapter 1 of \cite{NPLi11}).
\begin{figure}[h!]
\centering
\includegraphics[height=7.5cm,width=15cm]{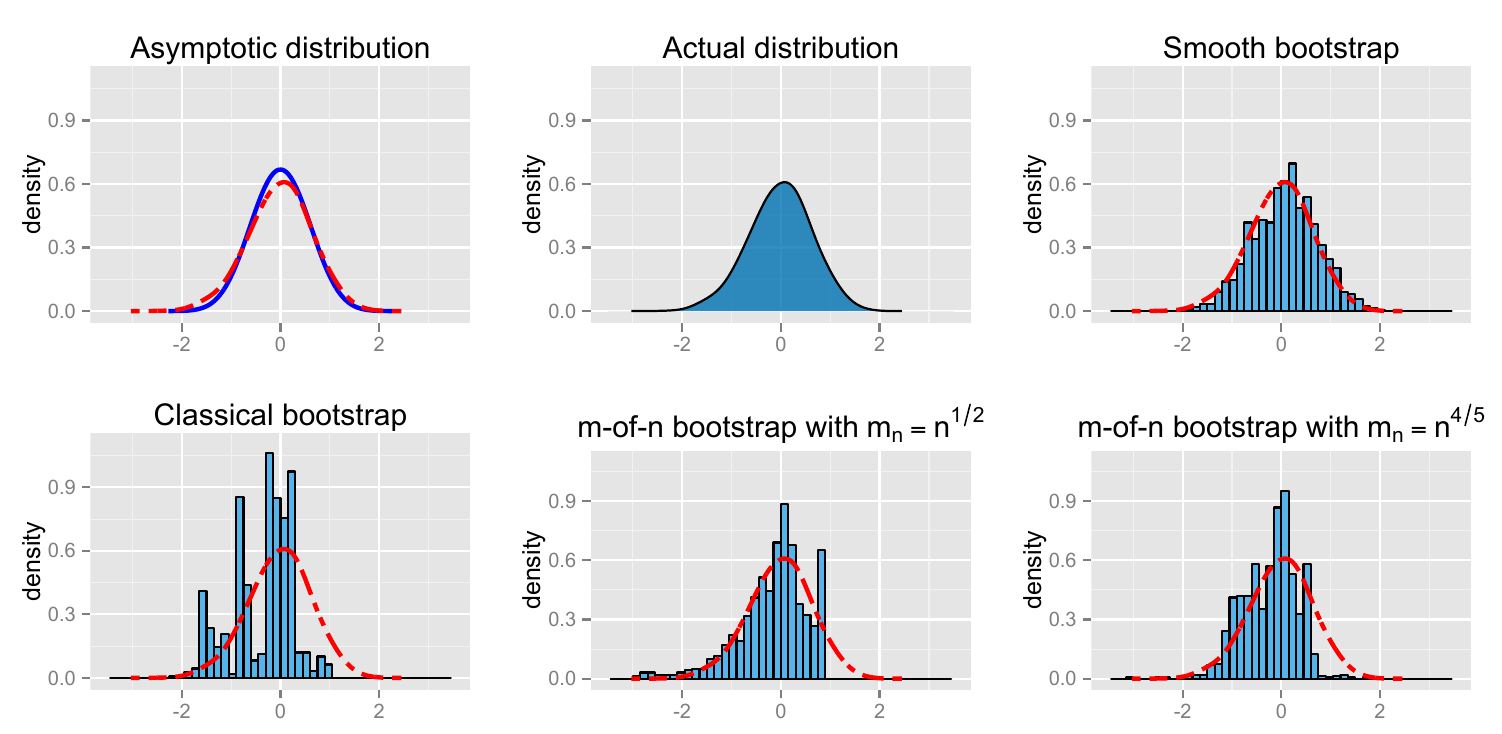}
\caption{Histogram of the distribution of $\Delta_{n,1}^*$  (i.e., the first coordinate of $\Delta^*_n$; see~\eqref{eq:DeltaStar}), conditional on the data, for four different bootstrap schemes. The density of $n^{1/3}(\hat{\beta}_{n,1}-\beta_{0,1})$ and its asymptotic limit are also plotted. The density of the asymptotic limit and the histograms are overlaid with the density of $n^{1/3}(\hat{\beta}_{n,1}-\beta_{0,1})$ (in dashed red).  Here $m_n$ denotes the bootstrap sample size for the two different $m$-out-of-$n$ bootstrap schemes.
}\label{figura2} 
\end{figure}

We next provide graphical evidence that illustrates the (non)-convergence of the different bootstrap schemes. We take $n=2000$ and  $d=2$ to construct histograms for the bootstrap approximation to the distribution of $n^{1/3}(\hat{\beta}_{n,1}-\beta_{0,1})$, obtained from $1000$ bootstrap samples for 4 different bootstrap schemes: the classical bootstrap, the smoothed bootstrap, and $m$-out-of-$n$ bootstrap with $m_n = \lceil\sqrt{n}\rceil$, $\lceil n^{4/5}\rceil$. 
In addition to all this, we give the histograms of the actual distribution of $n^{1/3}(\hat{\beta}_{n,1}-\beta_{0,1})$ and its asymptotic limit. For $\mathbb{\tilde P}$ described above, the asymptotic distribution of the first component of $n^{1/3}(\hat{\beta}_n-\beta_0)$ is that of $\frac{1}{\sqrt{2}}\argmax_{s\in\mathbb{R}}\Lambda(s)$ with $\displaystyle \Lambda(s):= 2^{-5/4}Z(s) - \frac{11}{30\sqrt{\pi}}s^2$, where $Z$ is a standard two-sided Brownian motion starting at 0. The resulting histograms are displayed in Figure \ref{figura2}.
\begin{center}
\begin{table}
\caption{The estimated coverage probabilities and average lengths of nominal 95\% confidence intervals for the first coordinate of $\beta_0$ obtained using the 5 different bootstrap schemes for each of the two models.}\label{table:sim}
{
\begin{tabular}{*{7}{c}}
\toprule
\multicolumn{7}{c}{$U|X\sim N(0,(1+|X|^2)^{-2}) ,\ X\sim \mbox{Uniform}([-1,1]^2),\ \beta_0 = 2^{-1/2}(1,1)^\top $}\\ 
\midrule
 &\multicolumn{2}{c}{$n=100$}  &\multicolumn{2}{c}{$n=200$}   &\multicolumn{2}{c}{$n=500$}  \bigstrut[t] \\
\cmidrule(r){2-3} \cmidrule(rl){4-5}  \cmidrule(l){6-7}
 & Coverage & Avg Length & Coverage & Avg Length & Coverage & Avg Length \\
\midrule
Classical & 0.68 & 0.91 & 0.75 & 0.58 & 0.75 & 0.49 \\ 
Smoothed & 0.79 & 0.67 & 0.89 & 0.53 & 0.93 & 0.41 \\ 
$\lceil n^{1/2} \rceil$ & 0.73 & 0.93 & 0.83 & 0.72 & 0.87 & 0.58  \\ 
$\lceil n^{2/3} \rceil$ & 0.72 & 0.89 & 0.83 & 0.70 & 0.87 & 0.56  \\ 
$\lceil n^{4/5} \rceil$ & 0.70 & 0.87 & 0.84 & 0.70 & 0.85 & 0.51  \\ 
\midrule
 &\multicolumn{2}{c}{$n=1000$}  &\multicolumn{2}{c}{$n=2000$}   &\multicolumn{2}{c}{$n=5000$}  \bigstrut[t] \\
\cmidrule(r){2-3} \cmidrule(rl){4-5}  \cmidrule(l){6-7}
 & Coverage & Avg Length & Coverage & Avg Length & Coverage & Avg Length \\
\midrule
Classical & 0.73 & 0.26 & 0.71 & 0.19 & 0.71 & 0.13 \\ 
Smoothed & 0.95 & 0.29 & 0.94 & 0.22 & 0.95 & 0.16 \\ 
$\lceil n^{1/2} \rceil$ & 0.91 & 0.41 & 0.89 & 0.33 & 0.97 & 0.23 \\ 
$\lceil n^{2/3} \rceil$ & 0.95 & 0.46 & 0.92 & 0.36 & 0.95 & 0.21 \\ 
$\lceil n^{4/5} \rceil$ & 0.89 & 0.34 & 0.86 & 0.24 & 0.89 & 0.16   \\
\bottomrule[\heavyrulewidth]
\end{tabular}
}
{
\begin{tabular}{*{7}{c}}
\multicolumn{7}{c}{$U|X\sim (1+|X|^2)^{-1}\Xi,\ \Xi\sim \mbox{Student}(3),\ X\sim \mbox{Uniform}([-1,1]^2),\ \beta_0 = 2^{-1/2}(1,1)^\top $}\\ 
\midrule
 &\multicolumn{2}{c}{$n=100$}  &\multicolumn{2}{c}{$n=200$}   &\multicolumn{2}{c}{$n=500$}  \bigstrut[t] \\
\cmidrule(r){2-3} \cmidrule(rl){4-5}  \cmidrule(l){6-7}
 & Coverage & Avg Length & Coverage & Avg Length & Coverage & Avg Length \\
\midrule
Classical & 0.66 & 0.82 & 0.70 & 0.73 & 0.74 & 0.41 \\
Smoothed & 0.78 & 0.66 & 0.89 & 0.53 & 0.93 & 0.40 \\ 
$\lceil n^{1/2} \rceil$ & 0.74 & 0.98 & 0.77 & 0.78 & 0.86 & 0.59 \\ 
$\lceil n^{2/3} \rceil$ & 0.73 & 0.96 & 0.77 & 0.74 & 0.87 & 0.58 \\ 
$\lceil n^{4/5} \rceil$ & 0.72 & 0.92 & 0.79 & 0.74 & 0.90 & 0.62 \\
\midrule
 &\multicolumn{2}{c}{$n=1000$}  &\multicolumn{2}{c}{$n=2000$}   &\multicolumn{2}{c}{$n=5000$}  \bigstrut[t] \\
\cmidrule(r){2-3} \cmidrule(rl){4-5}  \cmidrule(l){6-7}
 & Coverage & Avg Length & Coverage & Avg Length & Coverage & Avg Length \\
\midrule
Classical & 0.72 & 0.34 & 0.74 & 0.21 & 0.70 & 0.14 \\
Smoothed & 0.93 & 0.28 & 0.94 & 0.21 & 0.95 & 0.15 \\ 
$\lceil n^{1/2} \rceil$ & 0.87 & 0.50 & 0.92 & 0.36 & 0.93 & 0.27\\
$\lceil n^{2/3} \rceil$ & 0.92 & 0.55 & 0.94 & 0.40 & 0.97 & 0.26\\
$\lceil n^{4/5} \rceil$ & 0.88 & 0.49 & 0.88 & 0.29 & 0.87 & 0.17\\
\bottomrule
\end{tabular}
}
\end{table}
\end{center}

It is clear from Figure \ref{figura2} that the histogram obtained from the smoothed bootstrap (top-right) is the one that best approximates the actual distribution of $n^{1/3}(\hat{\beta}_{n,1}-\beta_{0,1})$ (top-center) and its asymptotic limit (top-left). Figure \ref{figura2} also illustrates the lack of convergence of the classical bootstrap as its histogram (bottom-right) is quite different from the ones in the top row. Although known to converge, the $m$-out-of-$n$ bootstrap schemes (bottom-left and bottom-center) give visibly asymmetric histograms with larger range, resulting in wider and more conservative confidence intervals.

We now study the performance of each of the bootstrap schemes by measuring the average length and coverage probability of the confidence intervals built from several random samples obtained from $\mathbb{\tilde P}$ for different choices of $d$. For $d=2,$ we simulate 1000 replicates of sample sizes $n=$ 100, 200, 500, 1000, 2000 and 5000. For each of these samples 5 different confidence intervals are built using the 4 bootstrap schemes discussed above and the $m$-out-of-$n$ bootstrap with $m_n =\lceil n^{2/3}\rceil$. In addition to considering $\mathbb{\tilde P}$ as the one used above, we conduct the same experiments with the following setting: $U|X\sim (1+|X|^2)^{-1}\Xi,\ \Xi\sim \mbox{Student}(3),\ X\sim \mbox{Uniform}([-1,1]^2),\ \beta_0 = 2^{-1/2}(1,1)^\top$ and $Y = \ind{\beta_0^\top X + U\geq 0} $, where Student(3) stands for a standard Student-$t$ distribution with 3 degrees of freedom. The results are reported in Table \ref{table:sim}.

Table \ref{table:sim} indicates that the smoothed bootstrap scheme outperforms all the others as it achieves the best combination of high coverage and small average length. Its average length is, overall, considerably smaller than those of the other consistent procedures. Needless to say, the classical bootstrap performs poorly compared to the others.


 To study the effect of dimension on the performance of the 5 different bootstrap schemes, we fix $n=10000$ and sample from $\mathbb{\tilde P}$ for $d=3,4 ,5,$ and $6.$ For each $d$, we consider 500 samples and for each sample we simulate 500 bootstrap replicates to construct the confidence intervals. The results are summarized in Table \ref{tab:dim}. An obvious conclusion of our simulation study is that the smoothed bootstrap is the best choice. 

It is easy to see from \eqref{ec1} that $\hat{\beta}_n$ is the maximizer of a step function which is not convex. Thus, the computational complexity of finding the maximum score estimator and that of bootstrap procedures increase  with sample size and dimension; see \cite{Manski86}, \cite{Pinkse93}, and \cite{Florios07} for discussions on the computational aspect of the maximum score estimator. All simulations in this paper were done on a  High Performance Computing (HPC) cluster with Intel E5-2650L processors  running R software over  Red Hat Enterprise Linux. For $d=6$ and $n=10000,$ each of the 500 independent replications took an average of 33 hours to evaluate the smoothed bootstrap confidence interval, while it took 23 hours to compute the classical bootstrap interval, and 3 hours  for the $m$-out-of-$n$ bootstrap procedure with $m_n=n^{4/5}$. We would like to point out that the routine implementing  the different bootstrap procedures  has not been optimized. Furthermore as bootstrap procedures are \textit{embarrassingly parallel},  distributed computing can be used to drastically reduce the computation time.  The following remarks are now in order.
\begin{center}
\begin{table}
\caption{The estimated coverage probabilities and average lengths (obtained from 500 independent replicates with 500 bootstrap replications)  of nominal 95\% confidence intervals for the first coordinate of $\beta_0$. For all values of $d,$ we have $n = 10^4$ i.i.d.~observations from $U|X\sim N(0,(1+|X|^2)^{-2}) ,\ X\sim \mbox{Uniform}([-1,1]^d), Y = \ind{\beta_0^\top X + U\geq 0},$ and $ \beta_0 = d^{-1/2}(1,\ldots,1)^\top \in\mathbb{R}^d.$}\label{tab:dim}
\begin{tabular}{*{9}{c}}
\toprule
 &\multicolumn{2}{c}{$d=3$}  &\multicolumn{2}{c}{$d=4$}   &\multicolumn{2}{c}{$d=5$}  &\multicolumn{2}{c}{$d=6$}  \bigstrut[t] \\
 \cmidrule(r){2-3} \cmidrule(rl){4-5}  \cmidrule(rl){6-7} \cmidrule(l){8-9} 
& \small{Coverage} & Len. & \small{Coverage} &  Len. & \small{Coverage} &  Len. & \small{Coverage} &  Len.\\
 \midrule
 Classical & 0.69 & 0.13 & 0.66 & 0.12 & 0.66 & 0.11 & 0.52 & 0.10  \\
Smoothed & 0.94 & 0.14 & 0.95 & 0.14 & 0.92 & 0.13 & 0.85 & 0.12\\
$\lceil n^{1/2} \rceil$ & 0.89 & 0.14 & 0.86 & 0.13 & 0.83 & 0.11& 0.74 & 0.11 \\
$\lceil n^{2/3} \rceil$ & 0.87 & 0.13 & 0.81 & 0.12 & 0.79 & 0.11& 0.67 & 0.10 \\
$\lceil n^{4/5} \rceil$ & 0.82 & 0.13 & 0.76 & 0.12 & 0.74 & 0.11& 0.61 & 0.10 \\
\bottomrule
\end{tabular}
\end{table}
\end{center}

\begin{remark}\label{rem:simul1}
 Given a bandwidth choice for $\hat{p}_n$ and $\hat{\kappa}_n$, a single bootstrap replicate for the classical and smoothed bootstrap procedure have the same computational complexity. However to evaluate the smoothed bootstrap confidence interval, we need to calculate  $\tilde{\beta}_n$ for the data set. This can be computationally intensive, especially when the sample size and dimension are large.  However, if we use the same kernel and  bandwidth choice  for both the kernel density estimator ($\hat{p}_n$) and the Nadaraya-Watson kernel regression estimator ($\hat{\kappa}_n$) then the computational complexity of evaluating $\tilde{\beta}_n$ can be greatly reduced. In all simulation examples of the paper, we have used the standard Gaussian kernel and bandwidth given by Scott's normal reference rule for both $\hat{p}_n$ and $\hat{\kappa}_n$. 
\end{remark}

\begin{remark}\label{rem:simul2}
We did not choose $m_n$ in the $m$-out-of-$n$ bootstrap method using any specific data-driven rule, as the computational complexities of such a method can be orders of magnitude higher. However, we tried $m_n = \lceil n^{1/3}\rceil$, $\lceil n^{9/10}\rceil$ and $\lceil n^{14/15}\rceil$ but their results were inferior to the reported choices ($\lceil n^{1/2}\rceil$, $\lceil n^{2/3}\rceil$ and $\lceil n^{4/5}\rceil$). Furthermore, we would like to point out that finite sample performance of smoothed bootstrap is superior to that of those obtained by \cite{deromi01} using subsampling (sampling without replacement) bootstrap  methods.
 \end{remark}

\section{A convergence theorem}\label{s3}
We now present a convergence theorem for triangular arrays of random variables arising from the binary choice model discussed above. This theorem will be used in Section~\ref{sec:mainproof} to prove Theorem \ref{t3}. 

Suppose that we are given a probability space $(\Omega,\mathcal{A},\mathbf{P})$ and a triangular array of random variables $\{(X_{n,j},Y_{n,j})\}^{n\in\mathbb{N}}_{1\leq j\leq m_n}$ where $(m_n)_{n=1}^\infty$ is a sequence of natural numbers satisfying $m_n \uparrow\infty$ as $n \rightarrow \infty$, and $X_{n,j}$ and $Y_{n,j}$ are $\mathbb{R}^d$ and $\{0,1\}$-valued random variables, respectively. Furthermore, assume that the rows {\small $\{(X_{n,1},Y_{n,1}),\ldots,(X_{n,m_n},Y_{n,m_n})\}$} are formed by i.i.d.~random variables.  We denote the distribution of $(X_{n,j},Y_{n,j})$, $1\leq j\leq m_n$, $n\in\mathbb{N}$, by  $\mathbb{Q}_{n}$ and the density of $X_{n,j}$ by $p_n.$ 
Recall the probability measure $\mathbb{P}$ on $\mathbb{R}^{d+1}$ and the notation introduced in Section~\ref{s2s1}. Denote by $\mathbb{P}_n^*$ the empirical measure defined by the row $(X_{n,1},Y_{n,1}), \ldots,(X_{n,m_n},Y_{n,m_n})$. Consider the class of functions 
\begin{align}
\mathcal{F}  := & \left\{f_{\alpha}(x,y):=\left(y - \frac{1}{2}\right)\ind{\alpha^\top x\geq 0} : \alpha \in\mathbb{R}^d \right\}, \label{eq:FnClassF} \\ 
\mathcal{G}  := & \left\{g_{\beta}(x):= \left(\kappa(x) - \frac{1}{2}\right)\ind{\beta^\top x\geq 0}:  \beta\in\mathbb{R}^d \right\}. \label{eq:FnClassG}
\end{align} 
We will say that $\beta_n^*\in\mathcal{S}^{d-1}$ is a {\it maximum score estimator} based on $(X_{n,i},Y_{n,i})$, $1\leq i\leq m_n$, if $$\beta_n^* = \argmax_{\beta\in\mathcal{S}^{d-1}} \frac{1}{m_n}\sum_{i=1}^{m_n} \left( Y_{n,i} - \frac{1}{2} \right)\ind{\beta^\top X_{n,i}\geq 0} =  \argmax_{\beta\in\mathcal{S}^{d-1}} \mathbb{P}_n^*(f_{\beta}),$$ where $f_\beta$ is defined in \eqref{eq:FnClassF}.  For any set Borel set $A \subset \R^d$, let $\nu_n(A):= \int_A p_n(x) dx.$ We take the measures $\{\mathbb{Q}_{n}\}_{n\in\mathbb{N}}$ and densities $\{p_n\}_{n \in \mathbb{N}}$  to satisfy the following conditions:

\begin{enumerate}[label=\bfseries (A\arabic*)]
\setcounter{enumi}{0}

\item 
 $\|\mathbb{Q}_n-\mathbb{P}\|_\mathcal{G}\rightarrow 0$ and the sequence $\{ \nu_n\}_{n=1}^\infty$ is uniformly tight. Moreover, $p_n$ is continuously differentiable on $\mathfrak{X}^\circ$ and $\nabla p_n$ is  integrable (with respect to the Lebesgue measure) over $\mathfrak{X}.$
\label{a1}

\item For each $n \in \mathbb{N}$ there is a continuously differentiable function $\kappa_n:\mathfrak{X}\rightarrow [0,1]$ such that 
\begin{equation*}\label{eq:kappa_n}
\kappa_n(x)=\mathbb{Q}_{n}(Y = 1|X=x)
\end{equation*}
 for all $n\in\mathbb{N}$, and $\|\kappa_n-\kappa\|_\texttt{X}\rightarrow 0$ for every compact set $\texttt{X}\subset\mathfrak{X}^\circ$.  \label{a2}
 
 \end{enumerate}
For $n\in\mathbb{N}$,  define $\Gamma_n:\mathbb{R}^d\rightarrow\mathbb{R}$ as
\begin{equation}\label{eq:DefGamma_n}
\Gamma_n(\beta):= \mathbb{Q}_n\left(f_{\beta}\right) = \int \left(\kappa_n(x) - \frac{1}{2} \right)\ind{\beta^\top x\geq 0}\; p_n(x) \,dx.
\end{equation}
 \begin{enumerate}[label=\bfseries (A\arabic*)]
\setcounter{enumi}{2}
\item Assume that   
\begin{equation}\label{eq:Def_beta_n} 
\beta_n := \argmax_{\beta \in\mathcal{S}^{d-1}} \Gamma_{n}(\beta),
\end{equation}
exists for all $n$, and 
\begin{equation}\label{surfint1}
\int_{\beta_n^\top  x=0}(\nabla\kappa_n(x)^\top \beta_n)\    p_n(x)xx^\top \ d\sigma_{\beta_n} \rightarrow \int_{\beta_0^\top  x=0}(\nabla\kappa(x)^\top \beta_0)\ p(x)xx^\top \ d\sigma_{\beta_0},
\end{equation}
where the above terms are standard surface integrals and $\sigma_{\beta_n}$ denotes the surface measure over $\{x\in\mathbb{R}^d : \beta_n^\top  x =0\}$, for all $n\geq 0$.  
\label{a3}
 \end{enumerate}


 Let $F_{n,K}$ be a measurable envelope of the class of functions $$\mathcal{F}_{n,K}:=\{\ind{\beta^\top x\geq 0}-\ind{\beta_n^\top  x\geq 0}: |\beta-\beta_n|\leq K\}.$$ Note that there are two small enough constants $C,K_*>0$ such that for any $0<K\leq K_*$ and $n\in\mathbb{N}$, $F_{n,K}$ can be taken to be of the form $\ind{\beta_K^\top x\geq 0 > \alpha_K^\top x} + \ind{\alpha_K^\top x\geq 0 > \beta_K^\top x}$ for $\alpha_K,\beta_K\in\mathbb{R}^d$ satisfying $|\alpha_K-\beta_K|\leq CK$. 
 \begin{enumerate}[label=\bfseries (A\arabic*)]
\setcounter{enumi}{3}
 
 \item  \label{a4} Assume that there exist $R_0,\Delta_0\in (0,K_* \land 1]$ and a decreasing sequence $\{\epsilon_n\}_{n=1}^\infty$ of positive numbers with $\epsilon_n\downarrow 0$ such that for any $n\in\mathbb{N}$ and for any $\Delta_0 m_n^{-1/3}<R\leq R_0$ we have
    \begin{enumerate} [label=(\roman*)]
    \item $\displaystyle |(\mathbb{Q}_n-\mathbb{P})(F_{n,R}^2)| \leq  \epsilon_1 R$; \label{a4-1}
    \item $\displaystyle \sup_{\begin{subarray}{c}|\alpha-\beta_n|\lor|\beta-\beta_n|\leq R\\ |\alpha-\beta|\leq R\end{subarray}}\big|(\mathbb{Q}_n-\mathbb{P})(\ind{\alpha^\top  X\geq 0}-\ind{\beta^\top  X\geq 0})\,\big| \leq  \epsilon_n R m_n^{-1/3}$. \label{a4-3}
    \end{enumerate}
\end{enumerate} 


In Lemma \ref{l3}, we show that  conditions \ref{a1}--\ref{a3} imply that $\Gamma_n$, as defined in \eqref{eq:DefGamma_n}, is twice continuously differentiable in a neighborhood of  $\beta_n$.  The main properties of $\Gamma$  and $\Gamma_n$ are established in Lemma \ref{l3} of the appendix.
\subsection{Consistency and rate of convergence} \label{s3s1}
In this sub-section we study the asymptotic properties of $\beta_n^*$.  Before attempting to prove any asymptotic results, we will state the following lemma, proved in Section~\ref{app:lemma_proof},  which establishes an important relationship between the $\beta_n$'s, defined in \eqref{eq:Def_beta_n}, and $\beta_0$.
\begin{lemma}\label{l1}
Under \ref{a1} and \ref{a2}, we have $\beta_n\rightarrow\beta_0$.
\end{lemma}
In the following lemma, proved in Section \ref{app:lemma_proof}, we show that $\beta_n^*$ is a consistent estimator of $\beta_0$.
\begin{lemma}\label{l2}
If  \ref{a1} and \ref{a2} hold, $\beta_n^*\cip\beta_0$.
\end{lemma}
We will now deduce the rate of convergence of $\beta_n^*$. It will be shown that $\beta_n^*$ converges  at rate $m_n^{-1/3}$. The proof of this fact relies on empirical processes arguments like those used to prove Lemma 4.1 in \cite{kipo}.  The following two lemmas, proved in Section \ref{app:lemma_proof}, adopt these ideas to our context (a triangular array in which variables in the same row are i.i.d.). The first lemma is a maximal inequality specially designed for this situation.

\begin{lemma}\label{l4}
Under \ref{a1}, \ref{a2}, and \ref{a4}, there is a constant $C_{R_0}>0$ such that for any $R>0$ and $n\in\mathbb{N}$ such that $\Delta_0m_n^{-1/3} \leq R m_n^{-1/3}\leq R_0$ we have
\[ \e{\sup_{|\beta_n-\beta|\leq R m_n^{-1/3}}\left\{|(\mathbb{P}_n^* - \mathbb{Q}_n)(f_{\beta} - f_{\beta_n})|\right\}^2} \leq C_{R_0}R m_n^{-4/3} \ \ \ \forall\ n\in\mathbb{N}.\]
\end{lemma}

With the aid of Lemma \ref{l4} we can now derive the rate of convergence of the maximum score estimator.

\begin{lemma}\label{l5}
Under \ref{a1}, \ref{a2}, and \ref{a4}, $m_n^{1/3}(\beta_n^* - \beta_n)=O_{\mathbf{P}}(1)$.
\end{lemma}

\subsection{Asymptotic distribution}\label{s3s2}
Before going into the derivation of the limit law of $\beta_n^*$, we need to introduce some further notation. Consider a sequence of matrices $(H_n)_{n=1}^\infty\subset\mathbb{R}^{d\times(d-1)}$ and $H\in\mathbb{R}^{d\times(d-1)}$ satisfying the following properties:
\begin{enumerate}[label=(\alph*)]
\item $\xi\mapsto H_n\xi$ and $\xi\mapsto H\xi$ are bijections from $\mathbb{R}^{d-1}$ to the hyperplanes $\{x\in\mathbb{R}^d: \beta_n^\top  x=0\}$ and $\{x\in\mathbb{R}^d: \beta_0^\top  x=0\}$, respectively.
\item The columns of $H_n$ and $H$ form orthonormal bases for $\{x\in\mathbb{R}^d: \beta_n^\top  x=0\}$ and $\{x\in\mathbb{R}^d: \beta_0^\top  x=0\}$, respectively.
\item There is a constant $C_H>0$, depending only on $H$, such that $\|H_n - H\|_2\leq C_H|\beta_n-\beta_0|$.
\end{enumerate}
We now give an intuitive argument for the existence of such a sequence of matrices. Imagine that we find an orthonormal basis $\{e_{0,1},\ldots,e_{0,d-1}\}$ for the hyperplane $\{x\in\mathbb{R}^d: \beta_0^\top  x=0\}$ and we let $H$ have these vectors as columns. We then obtain the {\it rigid motion} $\mathcal{T}:\mathbb{R}^d\rightarrow\mathbb{R}^d$ that moves $\beta_0$ to $\beta_n$ and the hyperplane $\{x\in\mathbb{R}^d: \beta_0^\top  x=0\}$ to $\{x\in\mathbb{R}^d: \beta_n^\top  x=0\}$. We let the columns of $H_n$ be given by $\{\mathcal{T}e_{0,1},\ldots,\mathcal{T}e_{0,d-1}\}$. The resulting sequence of matrices will satisfy the (a), (b) and (c) for some constant $C_H$. 

Note that (b) implies that $H_n^\top $ and $H^\top $ are the Moore-Penrose pseudo-inverses of $H_n$ and $H$, respectively. In particular, $H_n^\top H_n = H^\top  H = \mathcal{I}_{d-1}$, where $\mathcal{I}_{d-1}$ is the identity matrix in $\mathbb{R}^{d-1}$ (in the sequel we will always use this notation for identity matrices on Euclidean spaces). Additionally, it can be inferred from (b) that $H_n^\top (\mathcal{I}_d - \beta_n\beta_n^\top ) = H_n^\top $ and $H^\top (\mathcal{I}_d - \beta_0\beta_0^\top ) = H^\top $. Now, for each $s\in\mathbb{R}^{d-1}$ define 
\begin{equation}\label{eq:beta_ns}
\beta_{n,s}:= \left(\sqrt{1-(m_n^{-1/3}|s|)^2\land 1}\ \beta_n + m_n^{-1/3}H_n s\right)\ind{|s|\leq m_n^{1/3}} + |s|^{-1}H_n s\ind{|s|>m_n^{1/3}}.
\end{equation}
Note that $\beta_{n,s}\in\mathcal{S}^{d-1}$ as $\beta_n^\top H_n s=0$ and $|H_n s| =|s|$ for all $s \in \R^{d-1}$. Also, as $s$ varies in the set $|s| < m_n^{1/3}$, $\beta_{n,s}$ takes all values in the set $\{\beta \in \mathcal{S}^{d-1}: \beta_n^\top \beta > 0\}$. Furthermore, if $|s|\leq m_n^{1/3}$, $H_n s$ is the orthogonal projection of $\beta_{n,s}$ onto the hyperplane $\{x\in\mathbb{R}^d: \beta_n^\top  x=0\}$; otherwise $\beta_{n,s}$ is orthogonal to $\beta_n$. Define the process 
\begin{equation*}\label{eq:Lambda}
\Lambda_n(s):= m_n^{2/3}\mathbb{P}_n^*(f_{\beta_{n,s}} - f_{\beta_n}) 
\end{equation*}  and 
 \begin{equation*}\label{ecfin0}
s_n^* := \argmax_{s\in\mathbb{R}^{d-1}} \Lambda_n(s)  = \argmax_{s\in\mathbb{R}^{d-1}}\mathbb{P}_n^*(f_{\beta_{n,s}}).
\end{equation*}
Recall that $\beta_n^*=\argmax_{\beta \in \mathcal{S}^{d-1}}\mathbb{P}_n^*f_\beta$. As $\beta_n^*$ converges to $\beta_n$, $\beta_n^*$ will belong to the set $\{\beta \in \mathcal{S}^{d-1}: \beta_n^\top \beta > 0\}$ with probability tending to one. Thus, $\beta_{n,s_n^*}=\beta_n^*$  and $|s_n^*| < m_n^{1/3}$ with probability tending to one, as $n \to \infty$. Note that by \eqref{eq:beta_ns}, we have
$$ \beta_n^* = \sqrt{1-(m_n^{-1/3} s_n^*)^2\land 1}\, \beta_n+ m_n^{-1/3} H_n s^*_n,$$ when $|s_n^*| < m_n^{1/3}.$ Rearranging the terms in the above display, we get
$$ s_n^* =m_n^{1/3} H_n^\top(\beta_n^*-\beta_n) + m_n^{1/3} \Big[ \sqrt{1-(m_n^{-1/3}s_n^*)^2\land 1}-1\Big] H_n^\top\beta_n,$$ when $|s_n^*| < m_n^{1/3}$. As $H_n^\top\beta_n =0$ (from the defintion of $H_n$), we have
\begin{equation}\label{ecfin}
s_n^*  = m_n^{1/3} H_n^\top (\beta_n^* - \beta_n),
\end{equation}
with probability tending to 1, as $n\rightarrow \infty.$ Considering this, we will regard the processes $\{\Lambda_n\}_{n \ge 1}$ as random elements in the space of locally bounded real-valued functions on $\mathbb{R}^{d-1}$ (denoted by $\mathbf{B}_{loc}(\mathbb{R}^{d-1})$) and then derive the limit law of $s_n^*$ by applying the argmax continuous mapping theorem. We will take the space $\mathbf{B}_{loc}(\mathbb{R}^{d-1})$ with the topology of uniform convergence on compacta; our approach is based on that of \cite{kipo}.

To properly describe the asymptotic distribution we need to define the function $\Sigma:\mathbb{R}^{d-1}\times\mathbb{R}^{d-1}\rightarrow\mathbb{R}$ as follows:
\begin{align*}
\Sigma(s,t) :=& \frac{1}{4}\int_{\mathbb{R}^{d-1}} \{[(s^\top \xi)\land (t^\top \xi)]_+ + [(s^\top \xi)\lor (t^\top \xi)]_- \}p(H\xi)\ d\xi \nonumber\\
=& \frac{1}{8}\int_{\mathbb{R}^{d-1}}(|s^\top \xi| + |t^\top \xi| - |(s-t)^\top \xi|)p(H\xi)\ d\xi.\nonumber
\end{align*}
Additionally, denote by $W_n$ the $\mathbf{B}_{loc}(\mathbb{R}^{d-1})$-valued process given by \begin{equation*}
 W_n(s):= m_n^{2/3}(\mathbb{P}_n^*-\mathbb{Q}_n)(f_{\beta_{n,s}}-f_{\beta_n}).
\end{equation*}In what follows, the symbol $\rightsquigarrow$ will denote convergence in distribution. We are now in a position to state and prove our convergence theorem.

\begin{thm}\label{t1}
Assume that \ref{a1}--\ref{a4} hold. Then, there is a $\mathbf{B}_{loc}(\mathbb{R}^{d-1})$-valued stochastic process $\Lambda$ of the form $\Lambda(s) = W(s) + \frac{1}{2}s^\top H^\top \nabla^2\Gamma(\beta_0)Hs$, where $W$ is a zero-mean Gaussian process in $\mathbf{B}_{loc}(\mathbb{R}^{d-1})$ with continuous sample paths and covariance function $\Sigma$. Moreover, $\Lambda$ has a unique maximizer w.p.~1 and we have
\begin{enumerate}[label=(\roman*)]
\item $\Lambda_n\rightsquigarrow\Lambda$ in $\mathbf{B}_{loc}(\mathbb{R}^{d-1})$,
\item $\displaystyle s_n^*\rightsquigarrow s^*:= \argmax_{s\in\mathbb{R}^{d-1}} \Lambda(s) $,
\item $\displaystyle m_n^{1/3}(\beta_n^*-\beta_n)\rightsquigarrow Hs^*$.
\end{enumerate}
\end{thm}
\begin{proof}
Lemmas \ref{l8} and \ref{l9} imply that the sequence $(W_n)_{n=1}^\infty$ is stochastically equicontinuous and that its finite dimensional distributions converge to those of a zero-mean Gaussian process with covariance $\Sigma$. From Theorem 2.3 in \cite{kipo} we know that here exists a continuous process $W$ with these properties and such that $W_n\rightsquigarrow W$. By definition of $\Gamma_n$, note that  
 $\Lambda_n(\cdot)= W_n(\cdot)+m_n^{2/3}(\Gamma_n(\beta_{n,(\cdot)})-\Gamma_n(\beta_n))$. Moreover, from Lemma \ref{l3}, we have  {\small$m_n^{2/3}(\Gamma_n(\beta_{n,(\cdot)})-\Gamma_n(\beta_n))\cip \frac{1}{2}(\cdot)^\top H^\top \nabla^2\Gamma(\beta_0)H(\cdot)$} on $\mathbf{B}_{loc}(\mathbb{R}^{d-1})$ (with the topology of uniform convergence on compacta).  Thus, applying Slutsky's lemma (see e.g., Example 1.4.7, page 32 in~\cite{vw}) we get that $\Lambda_n\rightsquigarrow\Lambda$. The uniqueness of the maximizers of the sample paths of $\Lambda$ follows from Lemmas 2.5 and 2.6 in \cite{kipo}. Finally an application of Theorem 2.7 in \cite{kipo} gives $(ii)$, and $(iii)$ follows from \eqref{ecfin}.
\end{proof}

As a corollary we immediately obtain the asymptotic distribution of the maximum score estimator (taking $\kappa_n \equiv \kappa$ and $\beta_n \equiv \beta_0$) computed from i.i.d.~samples from $\mathbb{P}$.

\begin{cor}\label{c1}
If $(X_n,Y_n)_{n=1}^\infty\stackrel{i.i.d.}{\sim}\mathbb{P}$ and $\hat{\beta}_n$ is a maximum score estimator computed from $(X_i,Y_i)_{i=1}^n$, for every $n \ge 1$, then, 
\begin{equation*}
n^{1/3}(\hat{\beta}_n-\beta_0)\rightsquigarrow H \argmax_{s\in\mathbb{R}^{d-1}} \Lambda(s) .
\end{equation*}
\end{cor}
One final remark is to be made about the process $\Lambda$. The quadratic drift term can be rewritten, by using the matrix $H$ to evaluate the surface integral, to obtain the following more convenient expression
\[ \Lambda(s) = W(s) - \frac{1}{2}s^\top \left( \int_{\mathbb{R}^{d-1}}(\nabla\kappa(H\xi)^\top \beta_0)p(H\xi)\xi\xi^\top \ d\xi\right) s.\]

\noindent {\bf Remark:} Theorem~\ref{t1} gives us a general framework to prove the consistency of any bootstrap scheme. For example, if $\widehat{\mathbb{Q}}_n$ is an estimator of $\mathbb{P}$, computed from the data, such that \ref{a1}--\ref{a4} hold in probability or a.s.~(see the proof of Theorem~\ref{t3}), then the bootstrap scheme which generates bootstrap samples from $\widehat{\mathbb{Q}}_n$ will be consistent.

\section{Proof of Theorem \ref{t3}} \label{sec:mainproof}
In this section we use Theorem \ref{t1} to prove Theorem \ref{t3}.  We set $\kappa_n=\hat{\kappa}_n$, $\beta_n =\tilde{\beta}_n$.  
Let $\{n_k\}$ be a subsequence of $\mathbb{N}$. We will show that there exists a further subsequence $\{ n_{k_l}\}$ such that conditional on the data $\mathscr{Z}$, $\rho( \hat{\mathbb{G}}_{n_{k_l}}, \mathbb{G}) \stackrel{ a.s.}{\rightarrow} 0$. To show that $\rho( \hat{\mathbb{G}}_{n_{k_l}}, \mathbb{G}) \stackrel{}{\rightarrow} 0$ a.e.~data $(X_1,Y_1),\ldots,(X_n,Y_n)$ we appeal to Theorem~\ref{t1}. To apply Theorem~\ref{t1} we need to show that conditions \ref{a1}--\ref{a4} hold along the subsequence $\{ n_{k_l}\}$ for a.e.~$(X_1,Y_1),\ldots,(X_n,Y_n).$

First observe that \ref{smoth_2} implies that \ref{a2} holds along $\{ n_{k_l}\}$ a.s. We first show that $\|\widehat{\mathbb{Q}}_n-\mathbb{P}\|_\mathcal{G}\stackrel{P}{\rightarrow} 0$. Observe that by assumption \ref{smoth_1},
\[ |(\widehat{\mathbb{Q}}_n - \mathbb{P})(g_\beta(X))| = \left|\int_{  \mathfrak{X}} \left(\kappa(x) - \frac{1}{2}\right)\ind{\beta^\top x\geq 0} (\hat{p}_n(x) - p(x))\ dx\right| \leq \|\hat{p}_n - p\|_{  \mathfrak{X}} \cip  0.\]  
As $\hat p_n$ converges to $p$ uniformly, $\nu_n$ converges and is thus uniformly tight. This shows that \ref{a1} holds along    a further subsequence of $\{n_k\}$ for a.e.~data sequence.
Next we will show that~\ref{a3} hold in probability and the probability that inequalities \ref{a4} \ref{a4-1}--\ref{a4-3} holds tend to 1 as $n \to \infty$.

We will now show that \eqref{surfint1} holds in probability with $\kappa_n = \hat \kappa_n,$ $p_n=\hat{p}_n$ and $\beta_n = \tilde \beta_n$. The proof of \ref{a3} is slightly more involved and we describe the details below. Without loss of generality\footnote{For a general convex compact set $\mathfrak{X}$ with non-empty interior define $R^* := \sup \{R >0: \partial B(0,R) \cap \mathfrak{X} \ne \emptyset\}$, where $\partial B(0,R)$ is the boundary of the ball of radius $R$ around 0 in $\R^d$. For any $0 <\rho <R^*$, take $\texttt{X}_\rho := \mathfrak{X} \cap B(0,R^* - \rho)$. The proof would now follow with appropriate changes in the domains of integration.} we can assume that $\mathfrak{X}$ is the closed unit ball in $\R^d$ and write $\texttt{X}_\rho := (1 - \rho) \mathfrak{X}$, for any $0 < \rho <1$. By triangle inequality 
{\small \begin{align*}\label{eq:E1}
	 \left\| \int_{\tilde \beta_n^\top  x=0}(\nabla \hat \kappa_n(x)^\top \tilde \beta_n) \hat p_n(x)xx^\top d\sigma_{\tilde{\beta_n}} - \int_{\beta_0^\top  x=0}(\nabla\kappa(x)^\top \beta_0)p(x)xx^\top d\sigma_{\beta_0} \right\|_2
	\le U_n + Z_n+ V_n
\end{align*}}
where 
\begin{align*}
U_n  := & \left\| \int_{\tilde \beta_n^\top  x=0}(\nabla \hat \kappa_n(x)- \nabla\kappa(x))^\top \tilde\beta_n \hat p_n(x)xx^\top \ d\sigma_{\tilde \beta_n}\right\|_2, \\
Z_n:= & \left\| \int_{\tilde \beta_n^\top  x=0} \nabla\kappa(x)^\top \tilde \beta_n (\hat p_n(x)-p(x))xx^\top \ d\sigma_{\tilde \beta_n}\right\|_2,  \quad \text{and}\\ 
V_n  := & \left\| \int_{\tilde \beta_n^\top  x=0}(\nabla\kappa(x)^\top  \tilde \beta_n) \ p(x)xx^\top \ d\sigma_{\tilde \beta_n} - \int_{\beta_0^\top  x=0}(\nabla\kappa(x)^\top \beta_0)\ p(x)xx^\top \ d\sigma_{\beta_0} \right\|_2.
\end{align*}
Consider the matrices $H$ and $(H_n)_{n=1}^\infty$ described at the beginning of Section~\ref{s3s2}. Then $V_n$ can be expressed as 
{\small \begin{align*}
	&  \left\|\displaystyle\int_{\R^{d-1}}(\nabla \kappa (H_n \xi)^\top  \tilde \beta_n) \ p(H_n \xi) H_n \xi \xi^\top  H_n^\top  \ d\xi - \displaystyle\int_{\R^{d-1}}(\nabla\kappa(H \xi)^\top \beta_0)\ p(H \xi) H \xi \xi^\top  H^\top  \ d\xi \right\|_2.
\end{align*}}
As $p$ has compact support (both $\nabla \kappa$ and $p$ are bounded) and the $H_n$'s are linear isometries, we can apply the dominated convergence theorem to show that the above display goes to zero w.p.~1. Note that $\tilde \beta_n$ and thus $H_n$'s converge to $\beta_0$ and $H$, respectively, w.p.~1 as a consequence of Lemma~\ref{l1} and observation (c) in the beginning of Section~\ref{s3s2}. 

On the other hand, $U_n$ is bounded from above by 
\begin{align*}
\| \nabla \hat \kappa_n- \nabla\kappa \|_\mathfrak{X} \ \| \hat p_n \|_\mathfrak{X} \int_{\begin{subarray}{c} \tilde \beta_n^\top  x=0 \\ 1 - \rho \le |x| \le 1\end{subarray}} |x|^2  \ d\sigma_{\tilde \beta_n} + \| \nabla \hat \kappa_n- \nabla\kappa \|_{\texttt{X}_\rho} \ \|\hat p_n\|_\mathfrak{X} \int_{\begin{subarray}{c} \tilde \beta_n^\top  x=0 \\ |x| \le 1\end{subarray}} |x|^2 \ d\sigma_{\tilde \beta_n}.
\end{align*}
From the results in \cite{Blondin07} we know that $\| \nabla \hat \kappa_n(x)- \nabla\kappa(x) \|_{\texttt{X}_\rho} \stackrel{a.s.}{\rightarrow} 0$. Thus, noting that these surface integrals  are $\tilde \beta_n$-invariant, we get 
$$ U_n \le \| \nabla \hat \kappa_n - \nabla\kappa \|_\mathfrak{X}  \ \| \hat p_n \|_\mathfrak{X} \int_{\begin{subarray}{c} \beta_0^\top  x=0 \\ 1 - \rho \le |x| \le 1\end{subarray}} |x|^2  \ d\sigma_{\beta_0} + o_{\mathbf{P}}(1).$$

For any $\epsilon >0$ we can choose $M_\epsilon$ large enough, and $\rho_\epsilon$ sufficiently small so that 
{\small $$ \sup_{n \ge 1}  \mathbf{P}\Big(\| \nabla \hat \kappa_n- \nabla\kappa \|_\mathfrak{X} > M_\epsilon \Big) < \frac{\epsilon}{2}\; \; \mbox{ and } \;\;  M_\epsilon (\| p \|_\mathfrak{X}+o_p(n^{{-1}/{3}})) \int_{\begin{subarray}{c} \beta_0^\top  x=0 \\ 1 - \rho_\epsilon \le |x| \le 1\end{subarray}} |x|^2  \ d\sigma_{\beta_0}  < \frac{\epsilon}{2}. $$} Thus for all sufficiently large $n \in \mathbb{N}$, $$\mathbf{P}(U_n > \epsilon) \le \mathbf{P}(\| \nabla \hat \kappa_n- \nabla\kappa \|_\mathfrak{X} > M_\epsilon) + \mathbf{P}(U_n > \epsilon, \| \nabla \hat \kappa_n- \nabla\kappa \|_\mathfrak{X} \le M_\epsilon) < \epsilon.$$ 

Finally, $Z_n$ can be bounded from above as
\begin{align*}
&\left\| \int_{\tilde \beta_n^\top  x=0} \nabla\kappa(x)^\top \tilde \beta_n (\hat p_n(x)-p(x))xx^\top \ d\sigma_{\tilde \beta_n}\right\|_2 \\
\leq{}& \ \| \nabla\kappa\|_{\mathfrak{X}} \|\hat p_n-p\|_\mathfrak{X} \int_{\begin{subarray}{c} \tilde \beta_n^\top  x=0 \\ |x| \le 1\end{subarray}} |x|^2 \ d\sigma_{\tilde \beta_n} ={}o_p(n^{-1/3}).
\end{align*}
 To see that \ref{a4} holds, we will first show \ref{a4}-\ref{a4-3}. Observe that the set $\{ x \in \R^d: \alpha^\top x\geq 0> \beta^\top x\}$ is a multi-dimensional wedge-shaped region in $\R^d$, which subtends an angle of order $|\alpha-\beta|$ at the origin. As $\mathfrak{X}$ is a compact subset of $\R^d$ (assumption \ref{asum_c1}), we have that 
\begin{equation*}\label{eq:angular}
 \int_\mathfrak{X} |\ind{\alpha^\top x\geq 0} - \ind{\beta^\top x\geq 0}|\; dx \lesssim |\alpha-\beta|,
\end{equation*}
where by $\lesssim$ we mean bounded from above by a constant multiple; also see Example 6.4, page 214 of \cite{kipo}. Thus, for any $\alpha, \beta \in \mathbb{R}^d$, we have
\begin{align*}
 &|(\widehat{\mathbb{Q}}_n - \mathbb{P}) (\ind{\alpha^\top  X\geq 0}-\ind{\beta^\top  X\geq 0})| \\
 ={}&  \left|\int_{\mathfrak{X}} ( \ind{\alpha^\top x\geq 0} - \ind{\beta^\top x\geq 0})  (\hat{p}_n(x) - p(x))\ dx\right| \\
 \lesssim{}& |\alpha - \beta| \| \hat{p}_n - p \|_{\mathfrak{X}}.
\end{align*}
  It is now straightforward to show that \ref{a4}-\ref{a4-3}  will hold in probability because $\|\hat{p}_n - p\|_{\mathfrak{X}} = o_{\mathbf{P}}(n^{-1/3})$. A similar argument gives \ref{a4}-\ref{a4-1}.

\section{Proofs of results in  Section~\ref{s3s1}} 
\label{app:lemma_proof}
\subsection{Proof of Lemma \ref{l1}}
Let $\epsilon>0$ and consider a compact set $\texttt{X}_\epsilon$ such that $\mathbb{Q}_n(\texttt{X}_\epsilon\times\mathbb{R})>1-\epsilon$ for all $n\in\mathbb{N}$ (its existence is guaranteed by \ref{a1}). Then, $$|\Gamma_n(\beta) - \mathbb{Q}_n(g_\beta)|\leq 2\mathbb{Q}_n((\mathbb{R}^d\setminus\texttt{X}_\epsilon)\times\mathbb{R}) + \|\kappa_n-\kappa\|_{\texttt{X}_\epsilon}$$ for all $\beta\in\mathcal{S}^{d-1}$, where $g_\beta$ is defined in \eqref{eq:FnClassG}. 
 Consequently, \ref{a2}  shows that $\lsup \|\Gamma_n(\beta) - \mathbb{Q}_n(g_\beta)\|_{\mathcal{S}^{d-1}}\leq 2\epsilon$.  Moreover,  from \ref{a1}, we have that $\|\mathbb{Q}_n-\mathbb{P}\|_\mathcal{G}\rightarrow 0$. Therefore, as $\epsilon>0$ is arbitrary and $\Gamma (\beta) = \mathbb{P}(g_\beta)$, we  have \begin{equation} \label{eq:eq1l1}
\|\Gamma_n - \Gamma\|_{\mathcal{S}^{d-1}}\rightarrow 0. 
\end{equation}
 Considering that $\beta_0$ is the unique maximizer of the continuous function $\Gamma$ we can conclude the desired result as $\beta_n$ maximizes $\Gamma_n$ and the argmax function is continuous (under the sup-norm) for continuous functions on compact spaces with unique maximizers.

\subsection{Proof of Lemma \ref{l2}}
  Recall that $\beta_0$ is the well-separated unique maximizer of $\mathbb{P}(f_\beta) \equiv \Gamma(\beta).$ Thus, the result would follow as a simple consequence of the argmax continuous mapping theorem (see e.g., Corollary 3.2.3, \cite{vw}) if we can show that $ \|(\mathbb{P}_n^* -  \mathbb{P})(f_\beta)\|_{\mathcal{S}^{d-1}} \cip 0.$  
As $$\|(\mathbb{P}_n^* -  \mathbb{P})(f_\beta)\|_{\mathcal{S}^{d-1}}  \le \|(\mathbb{P}_n^* -  \mathbb{Q}_n)(f_\beta)\|_{\mathcal{S}^{d-1}} + \|\Gamma_n-\Gamma\|_{\mathcal{S}^{d-1}},$$ in view of \eqref{eq:eq1l1},  it is enough to show that $\e {\|\mathbb{P}_n^* -  \mathbb{Q}_n\|_{\mathcal{F}}}\rightarrow 0$. Now consider the classes of functions $\mathcal{F}_1:=\{y \ind{\alpha^\top x\geq 0}: \alpha \in\mathbb{R}^d\}$ and $\mathcal{F}_2:=\{\ind{\alpha^\top x\geq 0}: \alpha\in\mathbb{R}^d\}$. Note that  as $\mathcal{F} = \mathcal{F}_1 - \frac{1}{2}\mathcal{F}_2$, it follows that $\e {\|\mathbb{P}_n^* -  \mathbb{Q}_n\|_{\mathcal{F}}} \le \e {\|\mathbb{P}_n^* -  \mathbb{Q}_n\|_{\mathcal{F}_1}} + \e {\|\mathbb{P}_n^* -  \mathbb{Q}_n\|_{\mathcal{F}_2}}.$ Furthermore, observe that both $\mathcal{F}_1$ and $\mathcal{F}_2$ have the constant one as a measurable envelope function. The proof of the lemma would be complete if we can show the classes of functions $\mathcal{F}_1$ and $\mathcal{F}_2$ are manageable in the sense of definition 4.1 of \cite{poll89}, as by corollary 4.3 of \cite{poll89} we  will have  $\e {\|\mathbb{P}_n^* -  \mathbb{Q}_n\|_{\mathcal{F}_i}} \le J_i/\sqrt{m_n}$ for $i=1,2$, where the  constants $J_1$ and $J_2$ are  positive and finite. As VC-subgraph classes of functions with bounded envelope are manageable,  we will next show that  both $\mathcal{F}_1$ and $\mathcal{F}_2$  are  VC-subgraph classes of functions.   Since the class of all half-spaces of $\mathbb{R}^{d+1}$  is VC (see Exercise 14, page 152 in \cite{vw}), Lemma 2.6.18 in page 147 of \cite{vw} implies that both $\mathcal{F}_1 = \{y \ind{\alpha^\top x\geq 0}: \alpha\in\mathbb{R}^d\}$ and $\mathcal{F}_2$ are  VC-subgraph classes of functions. 

\subsection{Proof of Lemma \ref{l4}}
Take $R_0\leq K_*$, so for any $K\leq R_0$ the class $\{f_{\beta}-f_{\beta_n}\}_{|\beta-\beta_n|<K}$ is majorized by $F_{n,K}$. Our assumptions on $\mathbb{P}$ then imply that there is a constant $\tilde{C}$ such that $\mathbb{P}(F_{n,K}^2)=\mathbb{P}(F_{n,K})\leq \tilde{C} C K$ for $0<K\leq K_*$ (recall that $F_{n,K}$ is an indicator function). Note that the last inequality follows as $(\alpha,\beta)\mapsto\mathbb{P}(\beta^\top  X \leq 0<\alpha^\top  X)$ is continuously differentiable around $(\beta_n,\beta_n)$ (which can be shown using similar ideas as in Lemma \ref{l3}), and thus locally Lipschitz. Now, take $R>0$ and $n\in\mathbb{N}$ such that $\Delta_0 m_n^{-1/3}<R m_n^{-1/3}\leq R_0$. Since $\mathcal{F}_{n,R m_n^{-1/3}}$ is a VC-class (with VC index bounded by a constant independent of $n$ and $R$), the maximal inequality 7.10 in page 38 of \cite{poll90} implies the existence of a constant $J$, not depending neither on $m_n$ nor on $R$, such that $$\e{\|\mathbb{P}_n^* - \mathbb{Q}_n\|_{\mathcal{F}_{n,R m_n^{-1/3}}}^2}\leq J \mathbb{Q}_n(F_{n,R m_n^{-1/3}}) \ m_n^{-1}.$$ From \ref{a4}-\ref{a4-1} we can conclude that $$\e{\|\mathbb{P}_n^* - \mathbb{Q}_n\|_{\mathcal{F}_{n,R m_n^{-1/3}}}^2}\leq J (O(m_n^{-1/3})+ \tilde{C} C R m_n^{-1/3})\ m_n^{-1}$$ for all $R$ and $n$ for which $m_n^{-1/3} R \leq R_0$. This finishes the proof.

\subsection{Proof of Lemma \ref{l5}}
Take $R_0$ as in \ref{a4}, let $\epsilon>0$ and define
\begin{align*}
 M_{\epsilon,n} :=& \inf\left\{a>0: \sup_{\begin{subarray}{c}|\beta-\beta_n|\leq R_0\\ \beta\in\mathcal{S}^{d-1}\end{subarray}}\{|(\mathbb{P}_n^*-\mathbb{Q}_n)(f_{\beta} - f_{\beta_n})| - \epsilon|\beta-\beta_n|^2\}\leq  a m_n^{-2/3}\right\};\\
B_{n,j}:=& \{\beta\in\mathcal{S}^{d-1}: (j-1)m_n^{-1/3}<|\beta-\beta_n|\leq j m_n^{-1/3}\land R_0\}.
\end{align*}
Then, by Lemma \ref{l4} we have
\begin{align*}
 & \p{M_{\epsilon,n}> a} \\
={}& \p{\exists\ \beta\in \mathcal{S}^{d-1},\ |\beta-\beta_n|\leq R_0:\   |(\mathbb{P}_n^*-\mathbb{Q}_n)(f_{\beta} - f_{\beta_n})| > \epsilon|\beta-\beta_n|^2 + a^2 m_n^{-2/3}}\\
\leq{}& \sum_{j=1}^\infty \p{\exists\ \beta\in B_{n,j}: m_n^{2/3}|(\mathbb{P}_n^*-\mathbb{Q}_n)(f_{\beta} - f_{\beta_n})| > \epsilon^2(j-1)^2 + a^2 }\\
\leq{}& \sum_{j=1}^\infty \frac{m_n^{4/3}}{(\epsilon(j-1)^2 + a^2)^2}\e{\sup_{|\beta_n-\beta|< j m_n^{-1/3}\land R_0}\left\{|(\mathbb{P}_n^* - \mathbb{Q}_n)(f_{\beta}-f_{\beta_n})|\right\}^2}\\
\leq{}& C_{R_0}\sum_{j=1}^\infty \frac{j}{(\epsilon(j-1)^2 + a^2)^2} \rightarrow 0 \textrm{ as } a\rightarrow\infty.
\end{align*}
It follows that $M_{\epsilon,n}=O_\mathbf{P}(1)$. 
 From Lemma \ref{l3}-(c) we can find $N\in\mathbb{N}$ and $R,\epsilon>0$ such that $\Gamma_n(\beta) \le \Gamma_n(\beta_n) - 2\epsilon |\beta - \beta_n|^2$ for all $n\geq N$ and $\beta\in\mathcal{S}^{d-1}$ such that $0<|\beta - \beta_n|<R$. Since Lemma \ref{l2} implies $\beta_n^* - \beta_n\cip 0$, with probability tending to one we have
\begin{align*}
\mathbb{P}_n^*(f_{\beta_n^*}-f_{\beta_n}) \leq& \mathbb{Q}_n(f_{\beta_n^*}-f_{\beta_n}) + \epsilon |\beta_n^* - \beta_n|^2 + M_{\epsilon,n}^2m_n^{-2/3}\\
 \leq & \Gamma_n(\beta_n^*) - \Gamma_n(\beta_n) +\epsilon |\beta_n^* - \beta_n|^2 + M_{\epsilon,n}^2m_n^{-2/3}\\
 \leq& -\epsilon|\beta_n^* - \beta_n|^2 + M_{\epsilon,n}^2m_n^{-2/3}.
\end{align*}

Therefore, since $\beta_n^*$ is a maximum score estimator and $M_{\epsilon,n} = O_\mathbf{P}(1)$ we obtain that $\epsilon|\beta_n^* - \beta_n|^2 \leq \frac{3}{2}\epsilon_n |\beta_n^* - \beta_n|m_n^{-1/3} + O_\mathbf{P} (m_n^{-2/3})$. This finishes the proof.
\section{Acknowledgement} The third  author would like to thank Jason Abrevaya for suggesting him the problem.  We would also like to thank the Associate Editor and two anonymous referees for their helpful and constructive comments.

\appendix

\section{Auxiliary results for the proof of Theorem \ref{t1}}\label{app1}
\begin{lemma}\label{l3}
Denote by $\sigma_\beta$ the surface measure on the hyperplane $\{x\in\mathbb{R}^d:\beta^\top  x = 0\}$. For each $\alpha,\beta\in\mathbb{R}^d\setminus\{0\}$ define the matrix $A_{\alpha,\beta}:=(\mathcal{I}_d - |\beta|^{-2}\beta\beta^\top )(\mathcal{I}_d - |\alpha|^{-2}\alpha\alpha^\top ) + |\beta|^{-1}|\alpha|^{-1}\beta\alpha^\top.$ Note that, $x \mapsto A_{\alpha,\beta}x$ maps the region $\{\alpha^\top x \ge 0 \}$ to $\{\beta^\top x \geq 0\}$, taking $\{\alpha^\top x=0\}$ onto $\{\beta^\top x=0\}$. Recall the definitions of $\Gamma_n$ (see \eqref{eq:DefGamma_n}) and $\Gamma$ (see \eqref{eq:DefGamma}).  Then,
\begin{enumerate}[label=(\alph*)]
\item $\beta_0$ is the only maximizer of $\Gamma$ on $\mathcal{S}^{d-1}$ and we have
\begin{align*}
 \nabla\Gamma(\beta) =& \frac{\beta^\top \beta_0}{|\beta|^2}\left(\mathcal{I}_d-\frac{1}{|\beta|^2}\beta\beta^\top   \right)\int_{\beta_0^\top  x=0}\left(\kappa(A_{\beta_0,\beta} x)-\frac{1}{2}\right)p(A_{\beta_0,\beta} x)x\ d\sigma_{\beta_0},\\
 \nabla^2\Gamma(\beta_0) =& -\int_{\beta_0^\top  x=0}(\nabla\kappa(x)^\top \beta_0)\; p(x)xx^\top \ d\sigma_{\beta_0}.
\end{align*}
Furthermore, there is an open neighborhood $U\subset\mathbb{R}^d$ of $\beta_0$ such that {\small $\beta^\top \nabla^2\Gamma_0(\beta_0)\beta<0$} for all $\beta\in U\setminus \{t\beta_0:t\in\mathbb{R}\}$.
\item Under \ref{a1}--\ref{a3}, we have
\begin{align*}
 \nabla\Gamma_n(\beta) =& \frac{\beta^\top \beta_n}{|\beta|^2}\left(\mathcal{I}_d-\frac{1}{|\beta|^2}\beta\beta^\top   \right)\int_{\beta_n^\top  x=0}\left(\kappa_n(A_{\beta_n,\beta} x)-\frac{1}{2}\right)p_n(A_{\beta_n,\beta} x)x\ d\sigma_{\beta_n},\\
\nabla^2\Gamma_n(\beta_n) =&  -\int_{\beta_n^\top  x=0}(\nabla\kappa_n(x)^\top \beta_n)\;  p_n(x)xx^\top \ d\sigma_{\beta_n}.
\end{align*}
\item 
If conditions \ref{a1}--\ref{a3} hold, then  $\nabla^2\Gamma_n(\beta_n)\rightarrow\nabla^2\Gamma(\beta_0)$. Consequently, there is $N\geq 0$ and a subset $\widetilde{U}\subset U$ such that for any $n\geq N$, $\beta_n$ is a strict local maximizer of $\Gamma_n$ on $\mathcal{S}^{d-1}$ and $\beta^\top \nabla^2\Gamma_n(\beta_n)\beta<0$ for all $\beta\in\widetilde{U}\setminus \{t\beta_n:t\in\mathbb{R}\}$.
\end{enumerate}
\end{lemma}
\begin{proof}
We start with $(a)$. Lemma 2 in \cite{man85} implies that $\beta_0$ is the only minimizer of $\Gamma$ on $\mathcal{S}^{d-1}$. The computation of $\nabla\Gamma$ and $\nabla^2\Gamma$ are based on those in Example 6.4 in page 213 of \cite{kipo}. Note that for any $x$ with $\beta_0^\top  x = 0$ we have $\nabla\kappa(x)^\top \beta_0\geq 0$ (because for  $x$ orthogonal to $\beta_0$, $\kappa(x+ t\beta_0)\leq 1/2$ and $\kappa(x+ t\beta_0)\geq 1/2$ whenever $t<0$ and $t>0$, respectively). Additionally, for any $\beta\in\mathbb{R}^{d}$ we have:
\[ \beta^\top \nabla^2\Gamma(\beta_0)\beta = -\int_{\beta_0^\top  x=0}(\nabla\kappa(x)^\top \beta_0)(\beta^\top  x)^2p(x)\ d\sigma_{\beta_0}.\]
Thus, the fact that the set $\{x\in\mathfrak{X}^\circ: (\nabla\kappa(x)^\top \beta_0) p(x)>0\}$ is open (as $p$ and $\nabla\kappa$ are continuous) and intersects the hyperplane $\{x\in\mathbb{R}^d: \beta_0^\top  x = 0\}$ implies that $\nabla^2\Gamma(\beta_0)$ is negative definite on a set of the form $U\setminus \{t\beta_0:t\in\mathbb{R}\}$ with $U\subset\mathbb{R}^d$ being an open neighborhood of $\beta_0$.

We now prove $(b)$ and (c).  By conditions \ref{a1} and \ref{a2} we have that $\kappa_n$ is continuously differentiable on $\mathfrak{X}$ and $\nabla p_n$ is integrable on $\mathfrak{X}.$ Thus, we can compute $\nabla \Gamma_n$ by an application of the divergence theorem as in Example 6.4 in page 213 of \cite{kipo}. By the change of variable formula for measures (see Theorem 16.13, page 216 of \cite{bi3}), we can express $\nabla\Gamma_n(\beta)$ as {\small
\begin{equation*}
\beta_n^\top \beta_0\frac{\beta^\top \beta_n}{|\beta|^2}\left(\mathcal{I}_d-\frac{1}{|\beta|^2}\beta\beta^\top   \right)\int_{\beta_0^\top  x=0}\left(\kappa_n(A_{\beta_n,\beta}A_{\beta_0,\beta_n}x) -\frac{1}{2}\right)p_n(A_{\beta_n,\beta} A_{\beta_0,\beta_n}x)A_{\beta_0,\beta_n}x\ d\sigma_{\beta_0}.
\end{equation*}}
Starting with the above expression for $\nabla\Gamma_n$ we take the derivative with respect to $\beta$ using the product rule and differentiate under the integral sign. Recall that $\beta_n$ maximizes $\Gamma_n(\cdot)$, i.e., $\nabla\Gamma_n(\beta_n) = 0$. Thus, one of the terms in $\nabla^2\Gamma_n(\beta_n)$ will be zero as (note that $A_{\beta_n,\beta_n}=\mathcal{I}_d$) $$\int_{\beta_0^\top x=0}\left(\kappa_n(A_{\beta_0,\beta_n}x) -\frac{1}{2}\right)p_n( A_{\beta_0,\beta_n}x)A_{\beta_0,\beta_n}x\ d\sigma_{\beta_0}=0.$$ Hence, the only non-zero term in $\nabla^2\Gamma_n(\beta_n)$ is $-\int_{\beta_n^\top  x=0}(\nabla\kappa_n(x)^\top \beta_n) p_n(x)xx^\top d\sigma_{\beta_n}.$
Part (c) now follows immediately from (b) and  condition \ref{a3}.
\end{proof}
\begin{lemma}\label{l7}
Let $R>0$. Under \ref{a1}--\ref{a4} there is a sequence of random variables $\Delta_n^{R} = O_{\mathbf{P}}(1)$ such that for every $\delta>0$ and every $n\in\mathbb{N}$ we have,
$$\displaystyle \sup_{\begin{subarray}{c}|s-t|\leq\delta\\ |s|\lor|t|\leq R\end{subarray}}  \mathbb{P}_n^*\left[(f_{\beta_{n,s}} -f_{\beta_{n,t}})^2\right] \leq \delta \Delta_n^{R}m_n^{-1/3}.$$
\end{lemma}
\begin{proof}
Define $\mathcal{G}_{R,\delta}^n:=\{f_{\beta_{n,s}} - f_{\beta_{n,t}}:\ |s-t|\leq \delta,\ |s|\lor |t|\leq R\}$ and $\mathcal{G}_{R}^n:=\{f_{\beta_{n,s}}-f_{\beta_{n,t}}:\ |s|\lor |t|\leq R\}$. It can be shown that $\mathcal{G}_{R}^n$ is manageable with envelope $G_{n,R}:=3F_{n,2Rm_n^{-1/3}}$ (as $|\kappa_n -1/2| \leq 1$). Note that $G_{n,R}$ is independent of $\delta$.  Moreover, our assumptions on $\mathbb{P}$ then imply that there is a constant $\tilde{C}$ such that $\mathbb{P}(F_{n,K}^2)=\mathbb{P}(F_{n,K})\leq \tilde{C} C K$ for $0<K\leq K_*$, where the first equality is true because $F_{n,K}$ is an indicator function (see proof of Lemma \ref{l4} for more detail).  Considering this, condition \ref{a4}-\ref{a4-1} implies that
\begin{align*}
\mathbb{Q}_n  G^2_{n,R}\lesssim{}& \big| (\mathbb{Q}_n - \mathbb{P})(F^2_{n,2Rm_n^{-1/3}}) \big|+ \big|\mathbb{P}(F^2_{n,2Rm_n^{-1/3}})\big|\\
\lesssim{}& \epsilon_1  2Rm_n^{-1/3} + 2Rm_n^{-1/3}= O(m_n^{-1/3}).
\end{align*}  Thus, \ref{a4}-\ref{a4-3} and the maximal inequality 7.10 from \cite{poll90} show that there is a constant $\tilde{J}_{R}$ such that for all large enough $n$ we have
\begin{align*}
&\e{\sup_{\begin{subarray}{c}|s-t|\leq\delta\\ |s|\lor|t|\leq R\end{subarray}} \mathbb{P}_n^*\left[(f_{\beta_{n,s}}-f_{\beta_{n,t}})^2\right]} \\
\leq{}& 2 \e{\sup_{\begin{subarray}{c}|s-t|\leq\delta\\ |s|\lor|t|\leq R\end{subarray}} \mathbb{P}_n^*\left(|f_{\beta_{n,s}} - f_{\beta_{n,t}}|\right)}\\
\leq{}& 2\e{\sup_{f\in\mathcal{G}_{R}^n}  (\mathbb{P}_n^* - \mathbb{Q}_n)|f|} + 2 \sup_{f\in\mathcal{G}_{R,\delta}^n}\mathbb{Q}_n|f| \\
\leq{}& {m_n^{-1/2}} 4\epsilon_1\tilde{J}_{R}\sqrt{\mathbb{Q}_n G^2_{n,R}}+ 2 \sup_{f\in\mathcal{G}_{R}^n}\big| (\mathbb{Q}_n -\mathbb{P}) |f| \big|+ 2 \sup_{f\in\mathcal{G}_{R,\delta}^n}\mathbb{P} |f| \\
\leq{}& {m_n}^{-1/2}4\epsilon_1\tilde{J}_{R}\sqrt{O(m_n^{-1/3})} + \frac{2\epsilon_n R}{m_n^{1/3}} + 2\sup_{f\in\mathcal{G}_{R,\delta}^n} \mathbb{P}|f|.
\end{align*}
On the other hand, our assumptions on $\mathbb{P}$ imply that the function $\mathbb{P}(\ind{(\cdot)^\top  x \geq 0})$ is continuously differentiable, and hence Lipschitz, on $\mathcal{S}^{d-1}$. Thus, there is a constant $L$, independent of $\delta$, such that
\begin{equation*}\nonumber
\e{\sup_{\begin{subarray}{c}|s-t|\leq\delta\\ |s|\lor|t|\leq R\end{subarray}} \mathbb{P}_n^*\left[(f_{\beta_{n,s}}-f_{\beta_{n,t}})^2\right]} \leq o(m_n^{-1/3}) + \delta L m_n^{-1/3}.
\end{equation*}
The result now follows.
\end{proof}
\begin{lemma}\label{l8}
Under \ref{a1}--\ref{a4}, for every $R,\epsilon,\eta>0$ there is $\delta > 0$ such that
\[ \lsup_{n\rightarrow\infty}\p{\sup_{\begin{subarray}{c}|s-t|\leq\delta\\ |s|\lor|t|\leq R\end{subarray}}\left\{m_n^{2/3}\left|(\mathbb{P}_n^*-\mathbb{Q}_n) (f_{\beta_{n,s}}-f_{\beta_{n,t}})\right|\right\}>\eta} \leq \epsilon.
\]%
\end{lemma}
\begin{proof}
Let $\Psi_n := m_n^{1/3}\mathbb{P}_n^*(4F_{n,Rm_n^{-1/3}}^2) = m_n^{1/3}\mathbb{P}_n^*(F_{n,Rm_n^{-1/3}})$. Note that our assumptions on $\mathbb{P}$ then imply that there is a constant $\tilde{C}$ such that $\mathbb{P}(F_{n,K}^2)=\mathbb{P}(F_{n,K})\leq \tilde{C} C K$ for $0<K\leq K_*$ ($F_{n,K}$ is an indicator function). Considering this, conditions \ref{a4}-\ref{a4-1} and Lemma \ref{l4} imply that
\begin{align*}
\e{\Psi_n}  = & m_n^{1/3} \mathbb{Q}_n \left( \mathbb{P}_n^*F_{n,Rm_n^{-1/3}}\right)\\
 =& m_n^{1/3} \mathbb{Q}_n  F_{n,Rm_n^{-1/3}} \\ 
=& m_n^{1/3}(\mathbb{Q}_n - \mathbb{P})(F_{n,Rm_n^{-1/3}}) + m_n^{1/3}\mathbb{P}(F_{n,Rm_n^{-1/3}})= O(1).
\end{align*}
Now, define $$\Phi_n :=  m_n^{1/3}\sup_{\begin{subarray}{c}|s-t|\leq\delta\\ |s|\lor|t|\leq R\end{subarray}} \left\{ \mathbb{P}_n^*\left((f_{\beta_{n,s}}-f_{\beta_{n,t}})^2\right)\right\}.$$ The class of all differences $f_{\beta_{n,s}}-f_{\beta_{n,t}}$ with $|s|\lor|t|\leq R$ and $|s-t|<\delta$ is manageable (in the sense of definition 7.9 in page 38 of \cite{poll90}) for the envelope function $2F_{n,Rm_n^{-1/3}}$. By the maximal inequality 7.10 in \cite{poll90}, there is a continuous increasing function $J$ with $J(0)=0$ and $J(1)<\infty$ such that
\[ \e{\sup_{\begin{subarray}{c}|s-t|\leq\delta\\ |s|\lor|t|\leq R\end{subarray}}\left\{\left|(\mathbb{P}_n^*-\mathbb{Q}_n) (f_{\beta_{n,s}}-f_{\beta_{n,t}})\right|\right\}}\leq \frac{1}{m_n^{2/3}}\mathbb{Q}_n\left(\sqrt{\Psi_n}\function{J}{\Phi_n/\Psi_n}\right).\]
Let $\rho > 0$. Breaking the integral on the right on the events that $\Psi_n\leq \rho$ and $\Psi_n > \rho$ and the applying Cauchy-Schwartz inequality,
\begin{align*}
& \e{\sup_{\begin{subarray}{c}|s-t|\leq\delta\\ |s|\lor|t|\leq R\end{subarray}}\left\{m_n^{2/3}\left|(\mathbb{P}_n^*-\mathbb{Q}_n) (f_{\beta_{n,s},\beta_n}-f_{\beta_{n,t},\beta_n})\right|\right\}} \\
 \leq{}& \sqrt{\rho}J(1) + \sqrt{\e{\Psi_n\ind{\Psi_n>\rho}}}\sqrt{\e{\function{J}{1\land (\Phi_n/\rho)}}},\\
 \leq{}& \sqrt{\rho}J(1) + \sqrt{\e{\Psi_n\ind{\Psi_n>\rho}}}\sqrt{\e{\function{J}{1\land (\delta\Delta_n^{R}/\rho)}}},
\end{align*}
where $\Delta_n^{R} = O_{\mathbf{P}}(1)$ is as in Lemma \ref{l7}. It follows that for any given $R,\eta,\epsilon>0$ we can choose $\rho$ and $\delta$ small enough so that the results holds.
\end{proof}

\begin{lemma}\label{l9}
Let $s,t,s_1,\ldots,s_N\in\mathbb{R}^{d-1}$ and write $\Sigma_N\in\mathbb{R}^{N\times N}$ for the matrix given by $\Sigma_N:=(\Sigma(s_k,s_j))_{k,j}$. Then, under \ref{a1}--\ref{a4} we have
\begin{enumerate}[label=(\alph*)]
\item $m_n^{1/3}\mathbb{Q}_n(f_{\beta_{n,s}} - f_{\beta_{n}})\rightarrow 0$,
\item $m_n^{1/3}\function{\mathbb{Q}_n}{(f_{\beta_{n,s}} - f_{\beta_{n}})(f_{\beta_{n,t}} - f_{\beta_{n}})}\rightarrow \Sigma(s,t)$,
\item $(W_n(s_1),\ldots,W_n(s_N))^\top \rightsquigarrow \textrm{N}(0,\Sigma_N)$,
\end{enumerate}
where $\textrm{N}(0,\Sigma_N)$ denotes an $\mathbb{R}^{N}$-valued Gaussian random vector with mean 0 and covariance matrix $\Sigma_N$ and $\rightsquigarrow$ stands for weak convergence.
\end{lemma}
\begin{proof}
$(a)$ First note that for large enough $m_n$,  by \eqref{eq:beta_ns}, we have 
$$m_n^{1/3}|\beta_{n,s}-\beta_n| \le  \Big|\sqrt{m_n^{2/3}- s^2 }\,-m_n^{1/3}\Big|+ |s| \le 2 |s|,$$where the second inequality is true as $b-a\leq \sqrt{b^2-a^2}$, when $b\ge a \ge0$ and the fact that $|H_n s|=|s|.$ Moreover, we have that $|\beta_{n,s}-\beta_n|\rightarrow 0$ and $m_n\rightarrow \infty.$ Now as $\beta_n$ is the maximizer of $\Gamma_n$, observe that by \eqref{eq:DefGamma_n}, we have 
\begin{align*}
m_n^{1/3}\mathbb{Q}_n(f_{\beta_{n,s}} - f_{\beta_{n}}) &= m_n^{1/3} (\Gamma_n({\beta_{n,s}}) - \Gamma_n({\beta_{n}}))\\
&= m_n^{1/3} \big[ \nabla\Gamma_n(\beta_n)^\top (\beta_{n,s}-\beta_n) + O(|\beta_{n,s}-\beta_n|^2)\big]\\
&= O(m_n^{1/3} |\beta_{n,s}-\beta_n|^2)\\
&= O(|\beta_{n,s}-\beta_n|)=o(1).
\end{align*} 
$(b)$  First note that $(\ind{U+\beta_n^\top  X\geq 0}-1/2)^2\equiv 1/4$ and \[(\ind{\beta_{n,s}^\top x\geq0} - \ind{\beta_{n}^\top x\geq0})(\ind{\beta_{n,t}^\top x\geq0} - \ind{\beta_{n}^\top x\geq0}) = \ind{(\beta_{n,s}^\top x)\land(\beta_{n,t}^\top x)\geq 0 > \beta_n^\top x} + \ind{\beta_n^\top x\geq 0 > (\beta_{n,s}^\top x)\lor(\beta_{n,t}^\top x)}.\]
In view of these facts and condition \ref{a4}-\ref{a4-3}, we have 
\begin{align} \label{eq:varcov_1}
\begin{split}
& m_n^{1/3}\function{\mathbb{Q}_n}{(f_{\beta_{n,s}} - f_{\beta_{n}})(f_{\beta_{n,t}} - f_{\beta_{n}})}\\ 
={}& m_n^{1/3}\function{\mathbb{P}}{(f_{\beta_{n,s}} - f_{\beta_{n}})(f_{\beta_{n,t}} - f_{\beta_{n}})}+ o(1)\\
  ={}&\frac{m_n^{1/3}}{4}\mathbb{P} \left(\ind{(\beta_{n,s}^\top x)\land(\beta_{n,t}^\top x)\geq 0 > \beta_n^\top x} + \ind{\beta_n^\top x\geq 0 > (\beta_{n,s}^\top x)\lor(\beta_{n,t}^\top x)} \right) +o(1).
\end{split}
\end{align}
Now consider the transformations $T_n:\mathbb{R}^{d}\rightarrow\mathbb{R}^d$ given by $T_n(x):= (H_n^\top  x; \beta_n^\top  x)$, where $H_n^\top  x\in\mathbb{R}^{d-1}$ and $\beta_n^\top x\in\mathbb{R}$. Note that $T_n$ is an orthogonal transformation so $det(T_n)=\pm 1$ and for any $\xi\in\mathbb{R}^{d-1}$ and $\eta\in\mathbb{R}$ we have $T_n^{-1}(\xi;\eta)=H_n\xi + \eta\beta_n$. Under this transformation, observe that
\begin{align*}
C_{n,\xi} :={}&\big\{x \in \R^d :(\beta_{n,s}^\top x)\land(\beta_{n,t}^\top x)\geq 0 > \beta_n^\top x\big\}\\
 ={}& \left\{ (\xi;\eta) \in\mathbb{R}^{d-1}\times \mathbb{R}:  -m_n^{-1/3}\frac{s^\top \xi}{\sqrt{1-m_n^{-2/3}|s|^2}}\land \frac{t^\top \xi}{\sqrt{1-m_n^{-2/3}|t|^2}}\leq \eta < 0\right\}. 
 \end{align*}
 Similarly, we  have
 \begin{align*}
D_{n,\xi} :={}&\big\{x \in \R^d :\beta_n^\top x \geq 0>(\beta_{n,s}^\top x)\land(\beta_{n,t}^\top x)\big\}\\
 ={}&\Big\{ (\xi;\eta) \in\mathbb{R}^{d-1}\times \mathbb{R}: 0\leq\eta <-m_n^{-1/3}\frac{s^\top \xi}{\sqrt{1-m_n^{-2/3}|s|^2}}\lor \frac{t^\top \xi}{\sqrt{1-m_n^{-2/3}|t|^2}}\Big\}.
 \end{align*}
 Applying the above change of variable ($x \mapsto T_n(x) \equiv (\xi;\eta) $) and Fubini's theorem to \eqref{eq:varcov_1}, for all $n$ large enough,
\begin{align*}
& m_n^{1/3}\function{\mathbb{Q}_n}{(f_{\beta_{n,s}} - f_{\beta_{n}})(f_{\beta_{n,t}} - f_{\beta_{n}})} =m_n^{1/3}\iint \left(\ind{C_{n,\xi}} + \ind{D_{n,\xi}}\right)\; p(H_n\xi + \eta\beta_n)\ d\eta d\xi.
\end{align*}
With a further change of variable $w=m_n^{1/3} \eta$ and an application of the dominated convergence theorem we have
{\small \begin{align*}
m_n^{1/3}\function{\mathbb{Q}_n}{(f_{\beta_{n,s}} - f_{\beta_{n}})(f_{\beta_{n,t}} - f_{\beta_{n}})} &\rightarrow& \frac{1}{4}\int_{\mathbb{R}^{d-1}}\left((s^\top \xi\land t^\top \xi)_+ + (s^\top \xi\lor t^\top \xi)_-\right)p(H\xi)\ d\xi.
\end{align*}}
$(c)$ Define $\zeta_n:=(W_n(s_1),\ldots,W_n(s_N))^\top $, $\tilde{\zeta}_{n,k}$ to be the $N$-dimensional random vector whose $j$-entry is $m^{-1/3}(f_{\beta_{n,s_j}}(X_{n,k},Y_{n,k})-f_{\beta_{n}}(X_{n,k},Y_{n,k}))$, $\zeta_{n,k}:= \tilde{\zeta}_{n,k} - \e{\tilde{\zeta}_{n,k}}$ and
$$\rho_{n,k,j}:=\function{\mathbb{Q}_n}{(f_{\beta_{n,s_k}} - f_{\beta_{n}})(f_{\beta_{n,s_j}} - f_{\beta_{n}})} -\function{\mathbb{Q}_n}{(f_{\beta_{n,s_k}} - f_{\beta_{n}})}\function{\mathbb{Q}_n}{(f_{\beta_{n,s_j}} - f_{\beta_{n}})}.$$

We therefore have $\zeta_n = \sum_{k=1}^{m_n} \zeta_{n,k}$ and $\e{\zeta_{n,k}}=0$. Moreover, $(a)$ and $(b)$ imply that   $\sum_{k=1}^{m_n}\var{\zeta_{n,k}}=\sum_{k=1}^{m_n}\e{\zeta_{n,k}\zeta_{n,k}^\top }\rightarrow \Sigma_N$. Now, take $\theta\in\mathbb{R}^N$ and define $\alpha_{n,k}:=\theta^\top \zeta_{n,k}$. In the sequel we will denote by $\|\cdot\|_\infty$ the $\mathbb{L}_\infty$-norm on $\mathbb{R}^N$. The previous arguments imply that $\e{\alpha_{n,k}}=0$ and that $s_n^2:=\sum_{k=1}^{m_n}\var{\alpha_{n,k}} = \sum_{k=1}^{m_n}\theta^\top \var{\zeta_{n,k}}\theta\rightarrow \theta^\top \Sigma_N\theta$. Finally, note that for all $\epsilon>0$,
\begin{align*}
 \frac{1}{s_n}\sum_{l=1}^{m_n}\e{\alpha_{n,l}^2\ind{|\alpha_{n,l}|>\epsilon s_n}} \leq& \frac{N^2\|\theta\|_\infty^2 m_n^{-2/3}}{s_n}\sum_{l=1}^{m_n}\mathbb{Q}_{n}(|\alpha_{n,l}|>\epsilon s_n)\\
 \leq& \frac{N^2\|\theta\|_\infty^2 m_n^{-2/3}}{s_n^3 \epsilon^2}\sum_{1\leq k,j\leq N}\theta_k\theta_j m_n^{1/3}\rho_{n,k,j}\rightarrow 0.
\end{align*}
By the Lindeberg-Feller central limit theorem we can thus conclude that $\theta^\top \zeta_n = \sum_{j=1}^{m_n}\alpha_{n,j}\rightsquigarrow \textrm{N}(0,\theta^\top \Sigma_N\theta)$. Since $\theta\in\mathbb{R}^N$ was arbitrarily chosen, we can apply the Cramer-Wold device to conclude $(c)$.
\end{proof}

\section{The latent variable structure}\label{Disc}  In this section we discuss the latent variable structure of the binary response model and give some equivalent conditions on its existence, that might be of independent interest.
The median restriction $\med{U|X}=0$ on the {\it unobserved} variable $U$ implies that $\beta_0^\top  x \ge 0$ if and only if $\kappa(x) \ge 1/2$ for all $x \in \mathfrak{X}$; see~\cite{man75}. This condition can be re-written as $$\beta_0^\top  x\left(\kappa(x)-\frac{1}{2}\right)\geq 0$$ for all $x\in\mathfrak{X}$. Moreover, provided that the event $\left[\kappa(X) \in \{0,1/2,1\}\right]$ has probability 0, the above condition is also sufficient for the data to be represented with this latent variable structure. We make this statement precise in the following lemma.

\begin{lemma}\label{latentlemma}
Let $X$ be an random vector taking values in $\mathfrak{X}\subset\mathbb{R}^d$ and let $Y$ be a Bernoulli random variable defined on the same probability space $(\Omega,\mathcal{A},\mathbf{P})$. Write $\kappa(x):=\e{Y|X=x}$. Then:
\begin{enumerate}[label=(\roman*)]
\item If there are $\beta_0\in\mathcal{S}^{d-1}$ and a random variable $U$ such that $\med{U|X}=0$ and $Y=\ind{U+\beta_0^\top  X\geq 0}$, then $\beta_0^\top  x\left(\kappa(x) - {1/2}\right)\geq 0$ for all $x\in\mathfrak{X}$.
\item Conversely, assume the event $\left[\kappa(X) \in \{0,1/2,1\}\right]$ has probability 0 and that $\beta_0^\top  x (\kappa(x)- 1/2)\geq 0$ for all $x\in\mathfrak{X}^\circ$. Then, there is a probability measure $\mu$ on $\mathbb{R}^{d+1}$ such that if $(V,U)\sim\mu$, then $V\stackrel{\mathscr{D}}{=}X$, $\med{U|V}=0$ and $(X,Y)\stackrel{\mathscr{D}}{=}(V,W)$, where $W = \ind{U+\beta_0^\top V\geq 0}$ and $\stackrel{\mathscr{D}}{=}$ denotes equality in distribution.
\item Moreover, if $(\Omega,\mathcal{A},\mathbf{P})$ admits a continuous, symmetric random variable $Z$ with a strictly increasing distribution function that is independent of $X$, then $V$ in (ii) can be taken to be identically equal to $X$.
\end{enumerate}
\end{lemma}
\begin{proof}
The proof of $(i)$ follows from the arguments preceding the lemma; also see \cite{man75}. To prove $(ii)$ consider an $\mathfrak{X}$-valued random vector $V$ with the same distribution as $X$ and an independent random variable $Z$ with a continuous, symmetric and strictly increasing distribution function $\Psi$. Define $$ U:= \frac{\beta_0^\top  V}{\Psi^{-1}\left(\kappa(V)\right)}\; Z \; \ind{\kappa(V) \notin \{0,1/2,1\}}$$ and let $\mu$ to be the distribution of $(V,U)$. Then, letting $W = \ind{U+\beta_0^\top V\geq 0}$, for all $v$ with probability (w.p.) 1,
\begin{align*}
\mathbf{P}(W = 1|V = v) = \mathbf{P}(U \ge -\beta_0^\top v|V = v) = \mathbf{P}(Z \le \Psi^{-1}\left(\kappa(v) \right)|V = v) = \kappa(v),
\end{align*}
where we have used the fact that $\beta_0^\top  V/\Psi^{-1}\left(\kappa(V)\right) > 0$ w.p.~1 (since {\small $\beta_0^\top  x(\kappa(x)-1/2)\geq 0$} is equivalent to $\beta_0^\top  x \ \Psi^{-1}(\kappa(x))\geq 0$). Thus $(ii)$ follows. Under the assumptions of $(iii)$ note that we can take $V$ to be identically equal to $X$ in the above argument and result follows. 
\end{proof}

\bibliography{Referencias}
\bibliographystyle{apalike}

\end{document}